# Adaptive Block Compressive Sensing: towards a real-time and low-complexity implementation

**JOSEPH ZAMMIT[1], (Member, IEEE), IAN J. WASSELL[1]. (Member, IET)**
[1]Computer Laboratory, William Gates Building, 15 JJ Thomson Ave, Cambridge CB3 0FD, United Kingdom

Corresponding author: Joseph Zammit (e-mail: jz390@cl.cam.ac.uk).

The authors would like to thank and acknowledge funding from the Cambridge Trust, the Engineering and Physical Sciences Research Council Centre for Doctoral Training in Sensor Technologies and Applications (EP/L015889/1), and the Endeavour (Malta) Scholarships Scheme.

**ABSTRACT** Adaptive block-based compressive sensing (ABCS) algorithms are studied in the context of the practical realisation of compressive sensing on resource-constrained image and video sensing platforms that use single-pixel cameras, multi-pixel cameras or focal plane processing sensors. In this paper, we introduce two novel ABCS algorithms that are suitable for compressively sensing images or intra-coded video frames. Both use deterministic 2D-DCT dictionaries when sensing the images instead of random dictionaries. The first uses a low number of compressive measurements to compute the block boundary variation (BBV) around each image block, from which it estimates the number of 2D-DCT transform coefficients to measure from each block. The second uses a low number of DCT domain (DD) measurements to estimate the total number of transform coefficients to capture from each block. The two algorithms permit reconstruction in real time, averaging 8 ms and 26 ms for $256 \times 256$ and $512 \times 512$ greyscale images, respectively, using a simple inverse 2D-DCT operation without requiring GPU acceleration. Furthermore, we show that an iterative compressive sensing reconstruction algorithm (IDA), inspired by the denoising-based approximate message passing algorithm, can be used as a post-processing, quality enhancement technique. IDA trades off real-time operation to yield performance improvement over state-of-the-art GPU-assisted algorithms of 1.31 dB and 0.0152 in terms of PSNR and SSIM, respectively. It also exceeds the PSNR performance of a state-of-the-art deep neural network by 0.4 dB and SSIM by 0.0126.

## I. INTRODUCTION

OUR interest in low-power, autonomous image and video sensors, for example, for use in wireless sensor networks [1], has led us to explore compressive sensing (CS) as a means of reducing complexity and power requirements at the sensor. In these scenarios, a large number of limited energy and storage encoders are usually deployed under the control of a relatively more complex decoder [2]. CS has been touted as an excellent data acquisition technique because it departs from the classical digital signal processing paradigm of sampling at the Nyquist frequency, source coding, channel coding, and transmission. In CS, the aim is to push computational complexity to the decoder, leading to simple encoders but complex reconstruction at the decoder [3]. When applied to image acquisition, the disadvantages of current schemes are that they do not allow for real-time reconstruction of images or require specialized hardware, such as a graphical processing unit (GPU), which increases complexity such that they might not be feasible or economical in a wireless sensor network with a large number of nodes.

The theory of CS developed by Candès, Romberg and Tao [4] and Donoho [5], allows us to acquire a significantly lower number of measurements than classical theory suggests. The developed theory has been backed up by the development of the single-pixel camera for image and video signals [6]. CS theory has also been applied in multi-pixel cameras [7] and focal plane processing image sensors [8].

In CS, an unknown, sparse vector $\mathbf{f} \in \mathbb{R}^N$ is sensed by multiplying it with a measurement matrix $\mathbf{\Phi} \in \mathbb{R}^{M \times N}$. The $M$ resulting measurements comprise vector $\mathbf{y} \in \mathbb{R}^{M}$. In image and video compression, the signal of interest is sparse in a transform domain with transform $\mathbf{\Psi}$, such that $\mathbf{y} = \mathbf{\Phi\Psi x}$ forms a set of underdetermined linear equations. In this case, $\mathbf{A} = \mathbf{\Phi\Psi}$ is called the sensing matrix. CS theory shows that a sparse signal can be completely recovered from $M << N$ measurements, provided the sensing matrix satisfies some condition, such as the restricted isometry property (RIP) [9].

Reconstructing the image from the measurements requires the solution of $\mathbf{y} = \mathbf{Ax}$. The direct approach is to solve the underdetermined system using $L_0$ minimization, which is, however, NP-hard. $L_1$ and total variation (TV) minimization recast the problem as a linear programming (LP) problem, which can be practically solved using state-of-the-art LP solvers. This, in itself, is nontrivial, and a significant number of reconstruction techniques have been proposed to accelerate the solution, such as matching pursuit [10], Bayesian [11], approximate message passing (AMP) [12], and recently denoising AMP (D-AMP) [13] and neural networks [14].

In image and video applications, the difficulty in collecting and reconstructing measurements is compounded by the large dimensions of the images (or frames in video). Therefore, block-based CS (BCS) has been proposed to ease the problem [15], [16], [17]. The image is partitioned into $B \times B$ pixel blocks, where $B$ is significantly smaller than the height $H$ or width $W$ of the image. The size of the measurement matrix is substantially smaller and hence easier to store. Computing the measurements requires inner products of size $B \times B$ rather than $H \times W$ and hence much more energy efficient, and each block can be measured independently, for example, using a multi-pixel camera, thereby accelerating the measurement and reconstruction of the image. These benefits come at the expense of reconstruction quality, for example, as measured using the peak signal-to-noise ratio (PSNR) and structural similarity index measure (SSIM) [18].

Another problem with BCS is that whereas an image is always highly sparse in a transform domain, this is not always the case with the sub-image blocks. Since the sparsity varies significantly, either the compression ratio $C_R = M/N$ is compromised or some of the blocks are not reconstructed successfully when Gaussian measurement matrices are employed. This is not the case when the block measurement matrix comprises deterministic 2D-DCT basis functions in zigzag order. Although the ensuing measurement matrix is not RIP compliant, the compressed sensing measurements can always recover the image blocks, trivially in tens of milliseconds, using the inverse 2D-DCT, because the location of the transform coefficients is known. We refer to compressed sensing using low-pass, 2D-DCT matrices as the L-DCT-ZZ algorithm.

A number of authors have proposed adapting the number of measurements $m$ per block, depending on the sparsity of the sub-image blocks, to solve some of the problems of BCS and increase reconstruction quality. Their work is summarized below. The L-DCT-ZZ algorithm can also benefit from adapting $m$ block by block. Specifically, in this paper, we propose two adaptive algorithms to estimate sparsity - block boundary variation (BBV) and DCT domain (DD) - that can be applied to compressive image sensing.

### A. CONTRIBUTIONS

The main contribution in this paper is the development of two novel adaptive block compressive sensing (ABCS) algorithms that employ the deterministic partial, low-pass, 2D-DCT sensing matrix, namely, (i) adaptive linear block boundary variation estimation (AL-DCT-BBV), and (ii) adaptive linear 2D-DCT with DCT Domain estimation (AL-DCT-DD). These algorithms estimate block sparsity in the 2D-spatial and 2D-DCT transform domains and vary the number of compressive measurements in each block, respectively, as shown in figure 1. The number of measurements in each block is increased when the sparsity is low (where low sparsity means that the block has more non-zero transform coefficients in the 2D-DCT domain).

Extensive MATLAB simulations show that our algorithms achieve state-of-the-art performance amongst CPU reconstructed, real-time algorithms that can be reconstructed from measurements collected from single-pixel cameras, multi-pixel cameras and focal plane processing image sensors. These algorithms are reconstructed using one iteration of the inverse 2D DCT transform, in under 8 ms and 30 ms for $256 \times 256$ and $512 \times 512$ images, respectively. Another advantage is that the 2D-IDCT will always reconstruct an image block correctly. When CS is used to reconstruct blocks sensed using random measurement matrices, successful reconstruction is probabilistic, and the failure probability increases as the blocks become less sparse.

Our second contribution is the iterative denoising algorithm (IDA), derived from the denoising approximate message passing (D-AMP) algorithm [13]. IDA can be used as a post-processing step to real-time AL-DCT-BBV and AL-DCT-DD algorithms, as shown in figure 1, albeit with a large decoding time penalty. However, IDA reconstruction achieves state-of-the-art (SOTA) PSNR and SSIM results amongst ABCS results published in the literature.

### B. STRUCTURE OF THE PAPER

Section II discusses related work in the field of ABCS. Section III introduces CS theory and reconstruction algorithms. Non-adaptive deterministic measurement algorithms are introduced in section IV. The AL-DCT-BBV and AL-DCT-DD algorithms are introduced and analysed in section V, with the post-processing IDA algorithm presented in section VI. The algorithms are investigated empirically in section VII, first comparing them with other adaptive CS algorithms, then with two state-of-the-art CS algorithms proposed in the literature: the GPU-assisted CREAM [19] and the deep neural network (DNN) BCS-Net [20]. Section VIII draws conclusions from the study.

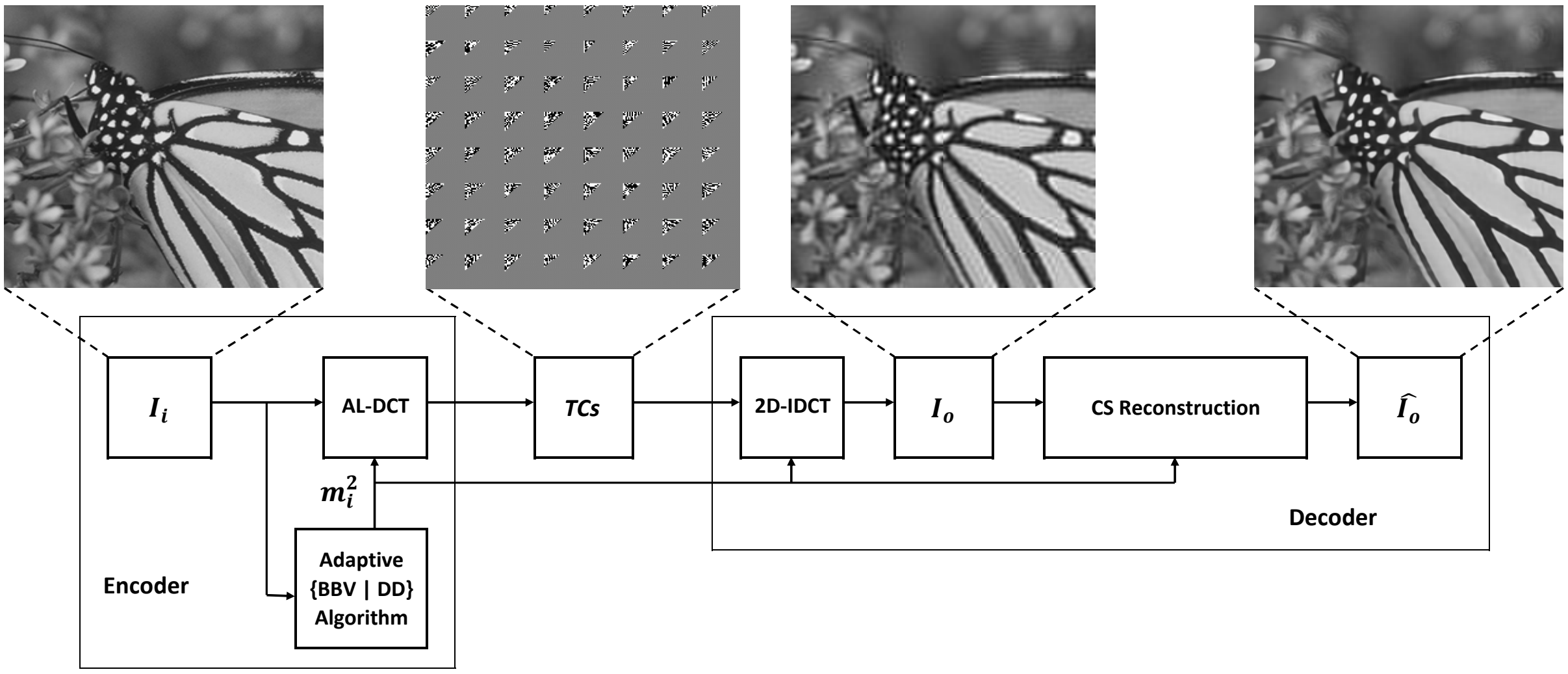

FIGURE 1. Adaptive AL-DCT-BBV and AL-DCT-DD system architecture. $I_i$ is the input image, $I_o$ is the image output in real time by the inverse 2D-DCT and $\widehat{I_o}$ is the higher quality image output by the CS reconstruction, for example, using the IDA algorithm. AL-DCT acquires $m_i^2$ adaptive CS measurements per block where each measurement is a 2D-DCT transform coefficient (TC).

### C. NOTATION USED IN THIS PAPER

Matrices are designated by bold capital letters, with vectors in bold lower case. Scalars are represented by normal letters, both upper and lower case, sometimes accompanied by subscripts. The superscript $i$ designates the value of vectors or scalars in the $i\ th$ iteration. Subscripts also designate the scalar components of vectors and matrices.

The non-adaptive linear L-DCT algorithm with low-pass transform coefficients collected in JPEG zigzag style is denoted L-DCT-ZZ. The adaptive versions of this algorithm are termed AL-DCT-BBV and AL-DCT-DD, with block boundary variation and DCT domain transform coefficient estimation, respectively. If the adaptive algorithms are further processed by the IDA reconstruction algorithm, we add the hyphenated suffixes -IDA-DnCNN or -IDA-BM3D to show whether the IDA algorithm uses the DnCNN [21] or BM3D [22] denoiser.

## II. RELATED WORK

Image BCS was first proposed by Lu Gan [15]. In this paper, the author proposed the development of spatially adaptive reconstruction algorithms as further work. Other seminal work on BCS of images and video was conducted by Fowler, Mun, Chen and Tramel [17], [16], [23], [24].

The non-adaptive, linear 2D-DCT CS technique, referred to here as L-DCT-ZZ, was inspired by the work of Romberg [25], who observed that CS of images is difficult and that a number of low-pass, linear 2D-DCT transform measurements, acquired in JPEG zigzag order [26], aid reconstruction significantly. In [25], Romberg compared low-pass, linear DCT acquisition in JPEG zigzag order (referred to in this paper as L-DCT-ZZ) with a scheme that combines some linear DCT measurements with noiselet measurements and concluded that the latter achieves better results.

Recently, Yuan and Haimi-Cohen showed that L-DCT-ZZ can be reconstructed using CS reconstruction techniques, referred to as the compressive sensing-based image compression system (CSbIC) in [27], and can achieve better SSIM results when compared with JPEG, especially at high compression ratios, though PSNR was reduced.

A significant body of literature has concentrated on the development of ABCS for images. Several authors estimate the number of measurements per block from measurements captured at the encoder. We refer to these as encoder-side (ES) techniques.

In some encoder-side papers, it is assumed that a full set of image pixels are already available to the encoder prior to compressive sensing and are used to perform the classification task. We refer to these as encoder-side full sensing (ES-FS) techniques. They are not feasible because in CS, it is assumed that capturing a few measurements from which we infer the pixel values is preferable to capturing the pixels themselves. We include these ES-FS schemes as benchmarks. Due to their inherent advantage in assuming that all pixels and/or transform coefficients are available before an image is adaptively sensed, they should achieve optimal results. A significant number of authors have published ES-FS techniques [28] [29] [30] [31] [32] [33] [34] and [35].

The BBV and DD algorithms proposed in this paper do not assume that a full set of pixels or transform measurements is available and are referred to as encoder-side compressive sensing (ES-CS) techniques. Other authors have proposed ES-CS schemes, such as [36] [37] [38] and [39].

Averbuch et al. proposed a departure from using random dictionaries that are incoherent with the sparse basis as used in classical CS [37]. Their dictionaries, instead, are nonrandom transform coefficient measurements from the sparsifying basis, for example, wavelets. They showed that these nonrandom measurements can still be captured on a DMD. This scheme is not block-based, however.

An alternative to ES-CS is to first transmit a reduced resolution image using normal BCS techniques, reconstruct it at the decoder, perform the analysis there, and feed back the subrate information to the encoder. We refer to these as decoder-side compressive sensing (DS-CS) techniques. [40] [36] and [41] proposed DS-CS ABCS techniques.

Several encoder-side papers ignore the measurements required to classify blocks prior to allocating the subrate to each block. Others include these measurements as part of the overall compressive measurements. Our BBV and DD algorithms include these estimation measurements as part of the block measurement allocation.

This paper uses SOTA reconstruction algorithms to increase PSNR and SSIM quality, with a reasonable reconstruction time. Donoho, Maleki and Montanari developed message passing algorithms for compressed sensing [12] with reduced reconstruction time. This led to the work on D-AMP by Metzler [13] [42], using block-matching and 3D filtering (BM3D) and DnCNN denoisers. The BM3D and DnCNN denoisers have also been used in recent work on image CS reconstruction using the plug-and-play method [43] [44] [45].

## III. COMPRESSIVE IMAGE SENSING AND RECONSTRUCTION

### A. COMPRESSIVE IMAGE SENSING

In sensing applications, it is often the case that information needs to be extracted from a set of measured data points. Taking $\mathbf{f} \in \mathbb{R}^N$ as the signal of interest and $\mathbf{y} \in \mathbb{R}^M$ as the measured signal, the relationship between the two can be written as:

$$\mathbf{y} = \mathbf{\Phi} \mathbf{f} \tag{1}$$

where $\mathbf{\Phi} \in \mathbb{R}^{M \times N}$ is the measurement matrix sampling signal $\mathbf{f}$ to give $\mathbf{y}$. If linear measurements of $\mathbf{f}$ are taken, then the reconstruction problem is reduced to solving a set of linear equations. The classical Nyquist-Shannon sampling theorem sets the sampling rate to twice the highest frequency for $\mathbf{f}$ to be reconstructed.

In a CS framework, the signal of interest $\mathbf{f}$ has a discrete representation $\mathbf{x} \in \mathbb{R}^N$ in some transform domain with basis $\mathbf{\Psi} \in \mathbb{R}^{N \times N}$ such that $\mathbf{f} = \mathbf{\Psi} \mathbf{x}$. Vector $\mathbf{x}$ is defined as sparse if it contains $S$ non-zero elements such that $S \ll N$. The measured signal $\mathbf{y}$ is therefore given by:

$$\mathbf{y} = \mathbf{\Phi} \mathbf{\Psi} \mathbf{x} \tag{2}$$

Grouping the measurement matrix and sparsifying basis matrix gives:

$$\mathbf{y} = \mathbf{A} \mathbf{x} \tag{3}$$

where $\mathbf{A} = \mathbf{\Phi} \mathbf{\Psi} \in \mathbb{R}^{M \times N}$ is defined as the sensing matrix. The goal in CS is thus to represent the original signal using $M < N$ samples, substantially below the Nyquist rate. As the locations of the $S$ sparse non-zero values of $\mathbf{x}$ are not known in advance, reconstruction involves solving the undetermined set of equations represented by (3) [46].

Two main questions arise from equation (3): (i) How should the measurement and sensing matrices be chosen; and (ii) how does one solve for $\mathbf{x}$? As the solution of (3) involves locating the non-zero components in vector $\mathbf{x}$, the most natural approach is an $L_0$ minimization, which can be written as:

$$\min_{\mathbf{x}} \|\mathbf{x}\|_0 \qquad s.t.\ \mathbf{A}\mathbf{x} = \mathbf{y} \tag{4}$$

As $L_0$ minimization is NP-hard and thus computationally intractable, a common approach is to rewrite the problem as an $L_1$ minimization:

$$\min_{\mathbf{x}} \|\mathbf{x}\|_1 \qquad s.t.\ \mathbf{A}\mathbf{x} = \mathbf{y} \tag{5}$$

Solving equation (5) can now be achieved via linear programming (LP), which is tractable [46]. This is commonly referred to as basis pursuit [46]. Faster reconstruction methods have been proposed in the literature, such as matching pursuit [10], approximate message passing (AMP) [12], denoising AMP (D-AMP) [13] and neural networks [14].

In real-world applications, noise is always present, and rewriting (3) to take this into account gives:

$$\mathbf{y} = \mathbf{A}\mathbf{x} + \mathbf{n} \tag{6}$$

where $\mathbf{n} \in \mathbb{R}^M$ represents Gaussian noise. The convex relation in (5) does not hold in this instance. However, it can be replaced by basis pursuit denoising (BPDN) [46] giving:

$$\min_{\mathbf{x}} \|\mathbf{x}\|_1 \qquad s.t.\ \|\mathbf{A}\mathbf{x} - \mathbf{y}\|_2^2 < \epsilon \tag{7}$$

where $\epsilon$ depends on $\mathbf{n}$.

An important feature of the sensing matrix is that it should retain the information pertinent to the sparse signal $\mathbf{x}$. There are several conditions that can determine whether a sensing matrix possesses this feature. Three well-known properties are (i) the restricted isometry property (RIP), (ii) the null space property, and (iii) mutual coherence. The sensing matrix that satisfies any of these three properties possesses the required information preservation feature [46].

It has been shown that when $\mathbf{A}$ is a random Gaussian matrix, it satisfies the RIP property [46] with high probability. Gaussian matrices are often used as the measurement matrices in the image CS literature. When the sensing matrix $\mathbf{\Phi}$ is Gaussian, the resulting sensing matrix $\mathbf{A} = \mathbf{\Phi}\mathbf{\Psi}$ is also Gaussian (because of the affine property of multivariate distributions) and benefits from satisfying the RIP with high probability. Simpler matrices, such as random (scrambled) Hadamard matrices, are also RIP compliant (studied as structured random sampling matrices in [46]) and have added implementation advantages, especially on DMD cameras [6]. However, optimal performance can be obtained using Gaussian or Bernoulli i.i.d matrices [4].

### B. RECONSTRUCTION ALGORITHMS

Many compressive sensing reconstruction algorithms exist to recover sparse signals from compressive samples. Recently, Pilastri and Tavares presented a taxonomy of reconstruction algorithms in compressive sensing [47]. In this taxonomy, the algorithms are grouped into six clusters: convex relaxation, greedy, non-convex, iterative, Bregman iterative, and combinatorial.

Reconstruction algorithms are characterized by their inherent complexity and the minimum number of compressive samples from which they can reliably recover the sparse coefficients and hence reconstruct the signal.

In image and video coding, we encounter signals whose transform in some domain is only approximately sparse; hence, we need reconstruction algorithms that can recover the predominant $S$ sparse components in the presence of "noise" consisting of a long tail of small, non-zero components. If the sorted coefficients $x_i$ decay with a power law, such that:

$$|x_i| \leq c \cdot i^{-p} \tag{8}$$

where $c$ is a constant and $p \geq 1$; then, $x_i$ is called $p$-compressible and can be approximated by a sparse signal (definition 2.2 in [46]). The transform coefficients of the 2D image and video data fit this model.

#### 1) Approximate Message Passing

In 2009, Donoho et al. [12] proposed approximate message passing (AMP) as a faster alternative to LP as a reconstruction technique for CS. AMP adds a correction term to iterative soft-thresholding (ISTA) [48] to vastly improve convergence.

In ISTA, the inverse noisy problem $\mathbf{y} = \mathbf{A}\mathbf{x} + \mathbf{n}$ is solved for $\mathbf{x}$ iteratively. Let $\mathbf{A}$ be the $M \times N$ sensing matrix with $M < N$, $\mathbf{y}$ a vector with $M$ measurements, and $\mathbf{x^t} \in \mathbb{R}^N$ the current estimate of the sparse solution. Then, starting from an initial guess $\mathbf{x^0} = 0$, $\mathbf{x^t}$ is computed by solving equations (9) and (10) iteratively:

$$\mathbf{x^{t+1}} = \eta_t(\mathbf{A}^*\mathbf{z^t} + \mathbf{x^t}) \tag{9}$$

$$\mathbf{z^t} = \mathbf{y} - \mathbf{A}\mathbf{x^t} \tag{10}$$

where $\mathbf{z^t}$ is the current error, $\mathbf{A}^*$ is the transpose of $\mathbf{A}$, and $\eta_t$ is a scalar threshold function that is applied component-wise. This simple iterative scheme can be applied to larger-scale applications than standard LP solvers because of the very lowcost per iteration and low storage requirements [12].

It is shown that adding the Onsager correction term $1/\delta\mathbf{z^{t-1}}\langle\eta'_{t-1}(\mathbf{A}^*\mathbf{z^{t-1}} + \mathbf{x^{t-1}})\rangle$ to (10), derived from belief propagation in graphical models, improves the convergence of the ISTA solution. The operator $\langle\mathbf{u}\rangle$ performs the component-wise mean of the vector $\mathbf{u}$, that is, $\langle\mathbf{u}\rangle \equiv \sum_{i=1}^{N} u_i/N$, $\eta'_t$ is the derivative of $\eta_t$ and $\delta = M/N$ (equivalent to $C_R$ in this paper).

The AMP algorithm then finds $\mathbf{x}$ iteratively by solving (11) and (12) below:

$$\mathbf{x^{t+1}} = \eta_{\mathbf{t}}(\mathbf{A}^*\mathbf{z^t} + \mathbf{x^t}) \tag{11}$$

$$\mathbf{z^t} = \mathbf{y} - \mathbf{A}\mathbf{x^t} + \frac{1}{\delta}\mathbf{z^{t-1}}\langle\eta'_{t-1}(\mathbf{A}^*\mathbf{z^{t-1}} + \mathbf{x^{t-1}})\rangle \tag{12}$$

The authors further show that the AMP solution has the same theoretical sparsity-undersampling trade-off as the LP-based reconstruction.

#### 2) Denoising AMP

Metzler et al. [13] extended the AMP framework by replacing the thresholding nonlinearity $\eta_t$ with a denoiser $D_\sigma$, which can be applied to reduce the Gaussian noise in the estimate $\mathbf{x_0} + \sigma\mathbf{z}$, where $\mathbf{z} = \mathbf{N(0, 1)}$ is Gaussian with zero mean and unit variance. They do this to leverage the rich literature on image denoisers, which has led to sophisticated algorithms that can remove Gaussian noise with large variance $\sigma^2$ from images. Their D-AMP iterations are given by equations (13) to (15):

$$\mathbf{x^{t+1}} = D_{\hat{\sigma}^t}(\mathbf{x^t} + \mathbf{A}^*\mathbf{z^t}) \tag{13}$$

$$\mathbf{z^t} = \mathbf{y} - \mathbf{A}\mathbf{x^t} + \frac{1}{\delta}\mathbf{z^{t-1}}\frac{div[D_{\hat{\sigma}^{t-1}}(\mathbf{x^{t-1}} + \mathbf{A}^*\mathbf{z^{t-1}})]}{m} \tag{14}$$

$$(\hat{\sigma}^t)^2 = \frac{\|\mathbf{z^t}\|_2^2}{m} \tag{15}$$

where $div[D(\mathbf{x})] = \sum_{i=1}^{n} \frac{\partial D(\mathbf{x})}{\partial x_i}$, $x_i$ is the $i^{th}$ element of $\mathbf{x}$, $n$ is the size of $\mathbf{x}$, and $m$ is the number of measurements.

The authors claimed that "D-AMP offered state-of-the-art CS recovery performance while operating tens of times faster than competing methods". Indeed, AMP combined with a BM3D [22] denoiser recovers a $128 \times 128$ pixel image, with close to state-of-the-art PSNR, in tens of seconds at a sampling rate of $10\%$. Other algorithms have since used the GPU to improve and accelerate reconstruction, such as CREAM [19] and BCS-Net [20]. D-AMP has also been upgraded to use the GPU-accelerated DnCNN denoiser by Metzler et al. [42].

## IV. IMAGE BCS USING DETERMINISTIC 2D-DCT SENSING MATRICES

In this paper, we focus on the use of deterministic, linear 2D transform sensing matrices such as 2D-DCT. The 2D discrete Walsh-Hadamard transform can also be used and has a simpler implementation because the measurement matrix consists of {+1/-1} elements. However, the PSNR and SSIM performance yielded by our simulations is inferior to 2D-DCT and is not reported here.

To accomplish compressive sensing, capitalizing on the energy compaction property of the 2D transform, $M$ low-pass transform coefficients are sensed and sorted in increasing order of sequency using the zigzag scanning procedure present in JPEG [26]. We refer to this as the L-DCT-ZZ algorithm.

If the objective is not compressive sensing but compression, all the transform coefficients may be collected (full sensing), and then only those exceeding some threshold are retained. It is possible to set one threshold for the whole image (THI) or one threshold per block (THB). In the first case, the number of retained transform coefficients per block varies block by block so that the THI algorithm is inherently adaptive. The number of retained transform coefficients in the THB algorithm is the same in each block, but the positions of the transform coefficients in the DCT domain are not known. The THI algorithm sets an upper bound to what can be achieved using adaptive 2D-DCT techniques. These two algorithms are defined formally in supplementary section III [49].

The advantage of deterministic sensing in L-DCT-ZZ is that the signals can be decoded very rapidly at the decoder, such that the collected images or frames can be used in real-time image and video transmission, for example, or to decrease the measurement time in magnetic resonance imaging (MRI). Block-based L-DCT-ZZ, similar to other block-based image compression algorithms, suffers from blocking effects at $C_R \leq 0.2$. Although the adaptive algorithms described below were found to attenuate blocking effects caused by the block-based measurements, some blocking is still visible when $C_R \leq 0.2$. Blocking can be removed in real time using the zigzag scan filter (ZZF) and thresholding filter (THF) described in supplementary section VI [49]. Better quality deblocking and image quality enhancement can be achieved using the CS reconstruction algorithms proposed and is studied in this paper, derived from Metzler's D-AMP algorithm [13].

## V. ADAPTIVE IMAGE BCS

The non-adaptive L-DCT-ZZ algorithm measures a fixed number of transform coefficients per block. In the adaptive schemes, the number of transform coefficients collected varies according to the sparsity of each block. Figure 1 shows a block diagram of the proposed adaptive system.

Two techniques for adapting the number of linear 2D-DCT transform coefficients in block compressive sensing are proposed, one in the spatial domain and the other in the 2D-DCT transform domain. Both techniques use two phases to encode an image adaptively. In the first phase, $m_i^1$ measurements are collected from each block $i$ and used to determine the number of additional transform coefficients, $m_i^2$, to be collected from all blocks in the second phase.

In the case of the DD technique, the $m_i^1$ measurements are collected as the DC and low-frequency components, as shown in figure 2, and determine the next $m_i^2$ measurements to be collected in phase 2.

All the measurements collected from both phases are considered for reconstruction purposes and hence the total number of measurements collected $M = \sum_{i=1}^{n_B}(m_i^1 + m_i^2)$, where $n_B$ is the number of blocks in the image.

The methods used to estimate $m_i^2$ must not be computationally intensive because they are implemented at the encoder side, and a target application for the methods described in this paper is for autonomously powered wireless image and video sensors. However, it is also possible to transmit the $m_i^1$ measurements to a decoder and use more elaborate algorithms at the decoder to estimate $m_i^2$ per block.

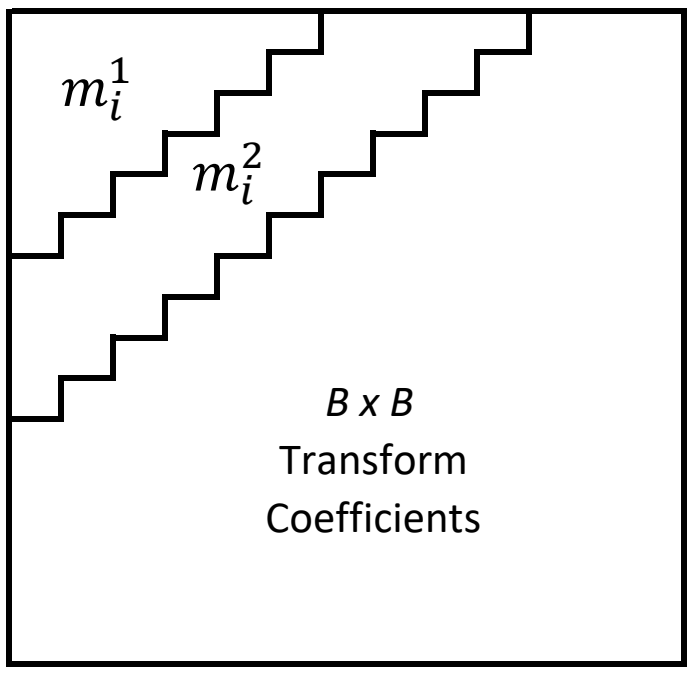

FIGURE 2. Adaptive L-DCT-DD phase 1 and 2 measurements.

This information is then fed back to the encoder as has been proposed in the literature [40], [36], [41]. In power-constrained applications, it is necessary to consider whether the power required to operate a receiver module to receive the $m_i^2$ feedback information is greater than the power required to just compute $m_i^2$ at the encoder. The requirement to wait for the reception of feedback information will also add to the reconstruction time of the image, although the increase may be contained if the sensing node is close to the receiver and the data rate of the link is sufficiently high.

Figure 1 shows the block AL-DCT measurements reconstructed in real time at the decoder using an inverse 2D-DCT transform and reconstituted into the output image $I_o$. The number of phase-two transform coefficients, $m_i^2$, is transmitted as side information. If better reconstruction quality is required, full-image CS reconstruction is invoked to output image $\widehat{I}_o$, albeit with a large reconstruction time penalty.

### A. ADAPTIVE L-DCT-ZZ USING BLOCK BOUNDARY VARIATION

The adaptive L-DCT-ZZ algorithm using BBV, AL-DCT-BBV, is described in Algorithm 1. The idea behind this technique is to adapt the number of transform coefficients collected in proportion to the total variation (TV) of pixel values around the perimeter of a block. It is inspired by the observation that if a block includes significant texture features, then these will likely cross-block boundaries and can be detected by measuring the TV at the block border. Since the number of pixels in the boundary is only $4B$, determining $m_i^2$ from just these pixels is more efficient than having to determine it from all $B^2$ pixels ($4B$ is the boundary constructed from efficient adjacent block measurements, as depicted in figure 3, it is not the $(4B-4)$ boundary of a block in isolation). When the compression factor $C_F = N/M$ in the block is high, the number of transform coefficients that can be collected is low, and the measurements required to calculate the BBV become a significant fraction of the whole budget for the block.

The measurement of the BBV can either be treated as a necessary overhead, or it is accounted for by reducing the number of transform coefficients that can be collected.

**Algorithm 1:** AL-DCT-BBV

**Input:** Image $\mathcal{I}$, compression factor $C_F$, block size $B$, Reconstruction Algorithm = {IDCT2|D-AMP|DAMP-D|IDA}.

**Output:** Image $\mathcal{O}$ sensed using AL-DCT-BBV and reconstructed using a reconstruction algorithm.

1 Crop and partition image, $\mathcal{I}$ into $n_B = \lfloor H/B \rfloor \cdot \lfloor W/B \rfloor$, $B \times B$ blocks;
2 Collect $m_i^1 = 2 \cdot n_S$ samples per block where $n_S = \lfloor B/C_F \rfloor$ measurements are equally spaced in the top and left-hand block border;
3 Estimate the number $m_i^2$ of additional samples to collect from each block $i$ where $m_i^2$ is given by equation (22);
4 Collect the $m_i^2$ transform coefficients from each block $i$;
5 Transmit $M = \sum_{i=1}^{n_B}(m_i^1 + m_i^2)$ measurements;
6 Reconstruct $\mathcal{O}$ from the M received samples using the selected reconstruction algorithm.

**Algorithm 2:** AL-DCT-DD

**Input:** Image $\mathcal{I}$, compression factor $C_F$, block size $B$, Reconstruction Algorithm = {IDCT2 | D-AMP|DAMP-D|IDA}}

**Output:** Image $\mathcal{O}$ sensed using AL-DCT-DD and reconstructed using reconstruction algorithm.

1 Crop and partition, $\mathcal{I}$ into $n_B = \lfloor H/B \rfloor \cdot \lfloor W/B \rfloor$, $B \times B$ blocks;
2 Calculate the number of non-adaptive measurements per block $m_i^1 = \lfloor B^2/(2 \cdot C_F) \rfloor$;
3 Collect $m_i^1$ L-DCT-ZZ coefficients from each block;
4 Estimate the number of L-DCT-ZZ transform coefficients to collect per block in phase 2, $m_i^2$, from equation (23);
5 Collect $m_i^2$ additional coefficients as required;
6 Transmit $M = \sum_{i=1}^{n_B}(m_i^1 + m_i^2)$ measurements;
7 Reconstruct $\mathcal{O}$ from the M received samples using the selected reconstruction algorithm.

This paper takes the latter view and attempts to minimize the number of measurements to estimate the BBV. This is accomplished in three ways. First, only the TV in the top row and left-hand column borders are measured in each block. Since the blocks stack to form an image, the right-hand TV is obtained from the next block to the right, and the bottom row TV is obtained from the bock below the current row. This reduces the number of pixels that need to be measured per block to $(2B-1)$. Second, rather than collecting pixel values, it is possible to collect pixel difference measurements, which involve only two adjacent pixels at a time. Thus, it is possible to reduce the measurements to approximately $B$ per block. Third, it is possible that not all adjacent pixel measurements are necessary and that the BBV can be estimated from fewer measurements. It is thus possible to collect a pixel difference measurement every $L$ pixels around the block border.

The right-hand blocks and the bottom blocks do not have right-hand and bottom neighbours that can contribute to the missing edges of the boundary. In these cases, the right-hand and bottom TVs can be measured from the current block.

Figure 3 shows BBV measurements in a $32 \times 32$ pixel block. The variation measurements in the current block, which consist of the absolute differences $|P_1 - P_2|$ of two adjacent pixels $P_1$ and $P_2$, are shown as rectangles outlined in black. $X_0$ is the pixel offset of the first variation measurement in a row, $Y_0$ is the pixel offset of the first measurement in a column, and $L$ is the stride between pixel variation measurements. In the figure, $X_0 = 1$, $Y_0 = 1$ and $L = 4$. We acquire phase 1 measurements that decrease as $C_F$ increases. Hence, we set $L$ empirically:

$$L = \lfloor C_F \rfloor \tag{16}$$

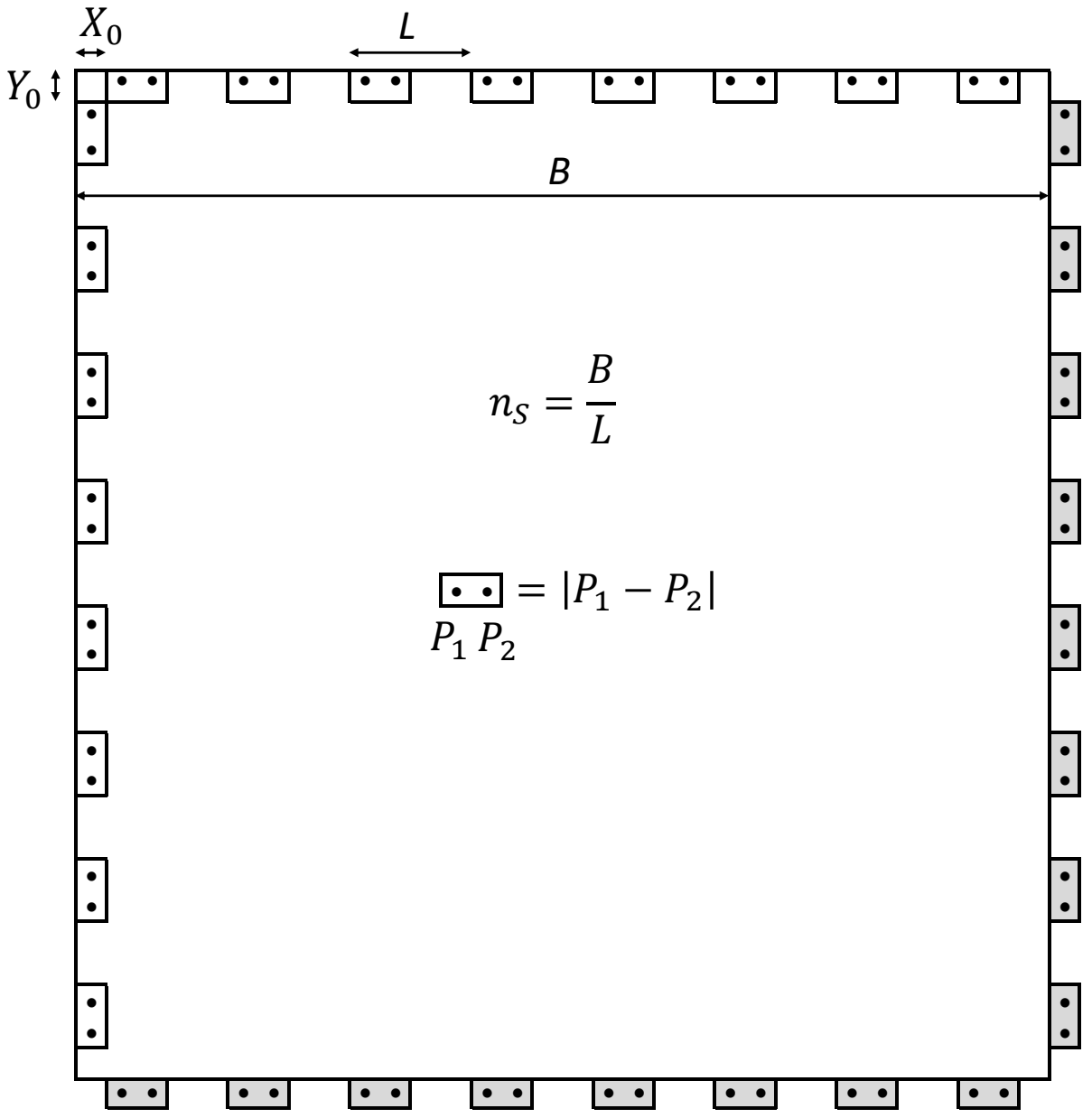

FIGURE 3. Block boundary variation measurements for a 32x32 block. The shaded measurements are imported from adjacent blocks.

and

$$X_0 = Y_0 = \lfloor L/2 \rfloor \tag{17}$$

The number of measurements $n_S$ per block side is given by:

$$n_S = \lfloor B/L \rfloor \tag{18}$$

and the number of measurements used to calculate the BBV is $4 \cdot n_S$, with half of them in the current block and the other half in adjacent blocks, shaded in grey in figure 3. $BBV_i$ is

given by:

$$BBV_i = \sum_{j=0}^{n_S-1} \{ |P_{(1,X_0+j\cdot L+1)} - P_{(1,X_0+j\cdot L)}| + |P_{(Y_0+j\cdot L+1,1)} - P_{(Y_0+j\cdot L,1)}| + |P_{(B+1,X_0+j\cdot L+1)} - P_{(B+1,X_0+j\cdot L)}| + |P_{(Y_0+j\cdot L+1),B+1)} - P_{(Y_0+j\cdot L,B+1)}| \} \tag{19}$$

If $C_F > B$, $n_s$ is zero according to equations (18) and (16). In this case, the algorithm reverts to being non-adaptive.

The number of blocks $n_B$ in an image of $H$ by $W$ pixels is given by:

$$n_B = \lfloor H/B \rfloor \cdot \lfloor W/B \rfloor \tag{20}$$

where the floor operation indicates that the image dimensions are cropped to be multiples of block size $B$. Then, the number of BBV measurements $M_{BBV}$ required in phase one is equal to the number of measurements in each block $2 \cdot n_S$ multiplied by the number of blocks $n_B$ and adding the measurements at the right and bottom edges and is given by:

$$M_{BBV} = 2 \cdot n_S \cdot n_B + (H + W)/L \tag{21}$$

These BBV measurements, in addition to being used to calculate the number of measurements $m_i^2$ required in phase two, can also be transmitted to the decoder to serve as additional reconstruction measurements. Then, assuming the total number of measurements used in image reconstruction is equal to $M$, the number of measurements in block $i$ during phase 2, $m_i^2$, is given by the number of remaining measurements $(M - M_{BBV})$ distributed amongst the blocks as a proportion of the block boundary variation measurement in block $BBV_i$ to the total block boundary measurements from all blocks and is hence given by:

$$m_i^2 = \left\lfloor (M - M_{BBV}) \cdot \frac{BBV_i}{\sum_{i=1}^{n_B} BBV_i} \right\rfloor \tag{22}$$

where $BBV_i$ is the BBV of block $i$, which is the sum of the absolute difference of two adjacent pixels at each measurement point around the boundary of the block. Half of the measurements are collected from block $i$, and the other half are collected from the adjacent blocks to the right and below, as shown in figure 3. $m_i^2$ has to be capped at $B^2 - m_i^1$, and any remaining measurements can be collected from other non-fully measured blocks.

### B. ADAPTIVE L-DCT-ZZ IN THE DCT DOMAIN

The adaptive L-DCT-ZZ algorithm in the DCT domain, AL-DCT-DD, described in Algorithm 2, uses $m_i^1$ linear 2D DCT measurements per block $i$, in zigzag scan order, to compute the number of measurements $m_i^2$ to collect in phase 2, as shown in figure 2. Several authors have considered first capturing a full set of measurements ($M = N$) to compute the value of all the pixels at the encoder and then adapting the number of measurements to transmit per block, as reported in section II. These full sensing techniques are feasible if the power required to sense the transform coefficients directly in the optical domain is less than the power required to capture pixels, convert them from analogue to digital, and then perform the transformation using digital computations. Zhu et al. [36] also consider a similar *significance-based allocation factor*, but this requires a whole block of transform coefficients. CS techniques, such as our AL-DCT-DD algorithm, collect fewer measurements ($M < N$) and hence require less sensing power than full sensing techniques.

The best PSNR performance by an adaptive algorithm in the DCT domain is achieved when using the full sensing THI algorithm that selects the highest valued transform coefficients from all the blocks. However, if compressive sensing is used, the strategy is to assign $m_i^2$ in proportion to the number $n_i$ of the $m_i^1$ measurements whose absolute value exceeds a threshold $T$. We found empirically that collecting half the number of available measurements in phase 1, that is, $n_B \cdot m_i^1 = M/2$, allows us to distribute a remaining equal number in proportion to $n_i$ such that $m_i^2$ is given by:

$$m_i^2 = \left\lfloor n_B \cdot m_i^1 \cdot \frac{n_i}{\sum_{i=1}^{n_B} n_i} \right\rfloor \tag{23}$$

The selection of $T$ is crucial to the performance of the algorithm. The best values were found empirically (refer to supplementary material, section IV [49]) and are tabulated in table 1 for different image heights $H$ and compression factors $C_F$.

| | **$H$ = 256** | **$H$ = 512** |
|---|---|---|
| **$T$ = 15** | $C_F \le 2$ | $C_F \le 2$ |
| **$T$ = 30** | $2 < C_F < 3.33$ | $2 < C_F < 10$ |
| **$T$ = 60** | $C_F \ge 3.33$ | $C_F \ge 10$ |

**TABLE 1.** Values of $T$ based on image height $H$ and $C_F$. The dynamic range of the pixel values is [0 255].

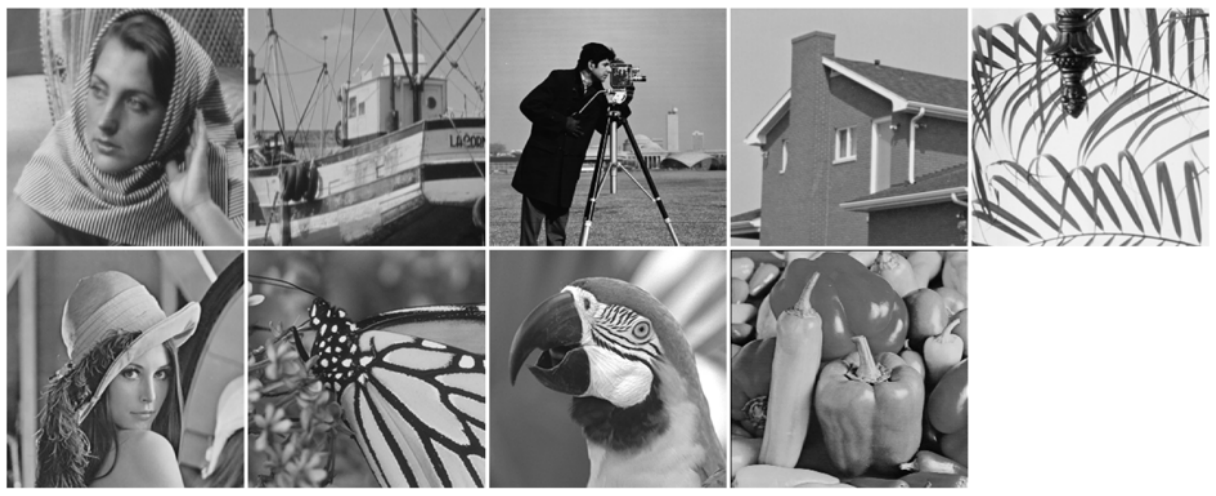

**FIGURE 4.** Set256: $256 \times 256$ image set.

### C. ANALYSIS OF THE BBV AND DD ADAPTIVE TECHNIQUES

Using the 2D DCT as the sparsifying domain, the number of transform coefficients in each block, $m_i$, that maximize PSNR, can be found by measuring all transform coefficients in all the blocks and selecting the M largest transform coefficients, irrespective of the blocks in which they occur. This follows from the orthogonal property of the DCT [50], meaning that it preserves the inner product. Since the sum of the square of the image pixels is equal to the sum of the square of the transform coefficients, if we select the highest valued transform coefficients, we are guaranteed to collect the highest signal energy.

| $\rho_{m_i,\widehat{m_i}}$ | Adaptive Algorithm = DD | | | | | | Adaptive Algorithm = BBV | | | | | |
|---|---|---|---|---|---|---|---|---|---|---|---|---|
| $C_R$ | 0.1 | 0.2 | 0.3 | 0.4 | 0.5 | Av | 0.1 | 0.2 | 0.3 | 0.4 | 0.5 | Av |
| 256 x 256 Image Set | 0.779 | 0.805 | 0.781 | 0.835 | 0.887 | 0.817 | 0.645 | 0.764 | 0.789 | 0.775 | 0.811 | 0.757 |
| 512 x 512 Image Set | 0.741 | 0.818 | 0.818 | 0.807 | 0.852 | 0.807 | 0.719 | 0.784 | 0.795 | 0.786 | 0.802 | 0.777 |

**TABLE 2.** Correlation coefficient $\rho_{m_i,\widehat{m_i}}$ for $256 \times 256$ and $512 \times 512$ image sets for both BBV and DD adaptive algorithms with varying $C_R$.

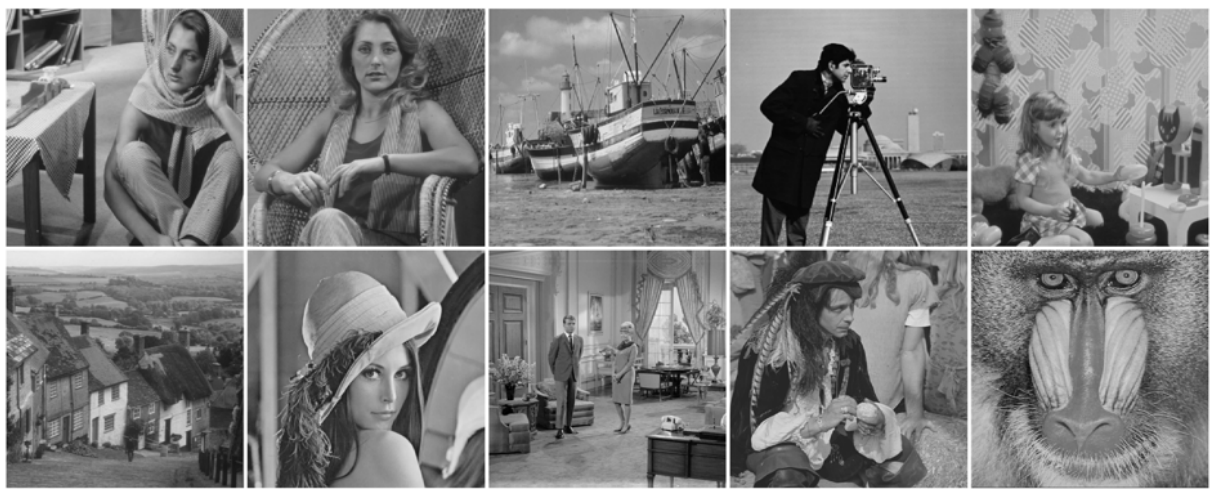

**FIGURE 5.** Set512: $512 \times 512$ image set.

The resulting algorithm is referred to as the THI algorithm in this paper. The higher the number of transform coefficients in a block $i$, the lower the sparsity in the block, and the more measurements that are required from that block. It is required to show that the number of measurements per block $\widehat{m_i}$ estimated by the BBV and DD algorithms are well correlated with the number of measurements $m_i$ that maximize PSNR.

The correlation coefficient $\rho_{m_i,\widehat{m_i}}$ and the confidence level $(1-p)$, where $p$ is the probability of the null hypothesis (that there is no correlation, i.e., $\rho_{m_i,\widehat{m_i}} = 0$), were measured for the Set256 and Set512 image sets in figures 4 and 5 for $C_R = \{0.1...0.5\}$. In all cases, the probability of the null hypothesis was always below 0.0001, indicating strong confidence in the correlation results. The average correlation results $\rho_{m_i,\widehat{m_i}}$ for the $256 \times 256$ and $512 \times 512$ image sets are tabulated in table 2 for both the BBV and DD algorithms (original image sets are available in [49]). The full results are reported in the supplementary material, section V [49].

For both image sets, the best correlation was obtained with the DD algorithm. The correlation was strongest at $C_R = 0.5$. The lowest correlation occurs for BBV at $C_R = 0.1$ and may be due to the low level of measurements acquired to compute $m_i$, since the number of block boundary measurements is proportional to $C_R$. This indicates that BBV should probably have a minimum number of boundary measurements from a correlation perspective, although this could impact PSNR and SSIM, because each phase one measurement reduces the number of DCT measurements.

## VI. RECONSTRUCTING DETERMINISTIC CS USING ITERATIVE DENOISING-BASED ALGORITHMS

In this section, we build on the work of Metzler [13] to design iterative algorithms that better reconstruct images compressively sensed using AL-DCT-BBV and AL-DCT-DD. The iterative denoising-based reconstruction algorithm is represented in block diagram form in figure 6. In D-AMP:

$$\alpha^t = \frac{1}{\delta}\frac{div[D_{\hat{\sigma}^{t-1}}(\mathbf{x^{t-1}} + \mathbf{A}^*\mathbf{z^{t-1}})]}{m} \tag{24}$$

so that $\alpha^t\mathbf{z^{t-1}}$ is the Onsager term. The $\Delta$ block introduces a delay of one iteration, $D_{\hat{\sigma}_t}$ is the denoiser block, and $\mathbf{r^t}$ is given by:

$$\mathbf{r^t} = \mathbf{y} - \mathbf{A}\mathbf{x^t} \tag{25}$$

The other variables are as defined for D-AMP above. To use D-AMP on deterministic, block-based CS, deterministic forward transform is cast as a sensing matrix $\mathbf{A}$ which has a sparse representation:

$$\mathbf{A} = \begin{bmatrix} \mathbf{B}_{1,1} & & & & \\ & \ddots & & & \\ & & \mathbf{B}_{i,i} & & \\ & & & \ddots & \\ & & & & \mathbf{B}_{n_B,n_B} \end{bmatrix} \tag{26}$$

where $\mathbf{B}_{i,i}$ are $m \times n$ basis matrices with $n = B \times B$, $m = \lfloor n \times C_R \rceil$, $C_R$ is the compression ratio $M/N$, $B$ is the size of one side of the image block and $n_B$ is the number of blocks in the image. In non-adaptive CS, $m$ and $n$ are constant since there are equal measurements per block. In adaptive CS, $m$ varies per block. $\mathbf{A}^*$ is the transpose of $\mathbf{A}$.

When $\mathbf{B}$ is a random Gaussian block matrix, we empirically found that D-AMP frequently failed to recover some of the blocks correctly, leading to very significant blocking effects. This never happens when the row vectors of $\mathbf{B}$ are taken from the 2D-DCT basis functions, in zigzag order, with the first basis function being that for the DC component.

### 1) Modified D-AMP - DAMP-D

D-AMP was derived based on a Gaussian distribution of the residual reconstruction error at each iteration [13]. When the measurement matrix is a low-pass deterministic 2D-DCT matrix, this assumption no longer holds, and the reconstruction quality deteriorates. Figure 6 shows that the Onsager term implements an adaptive integration of the reconstruction error, akin to integral control in a closed-loop feedback system. This inspires us to vary the integral control loop gain

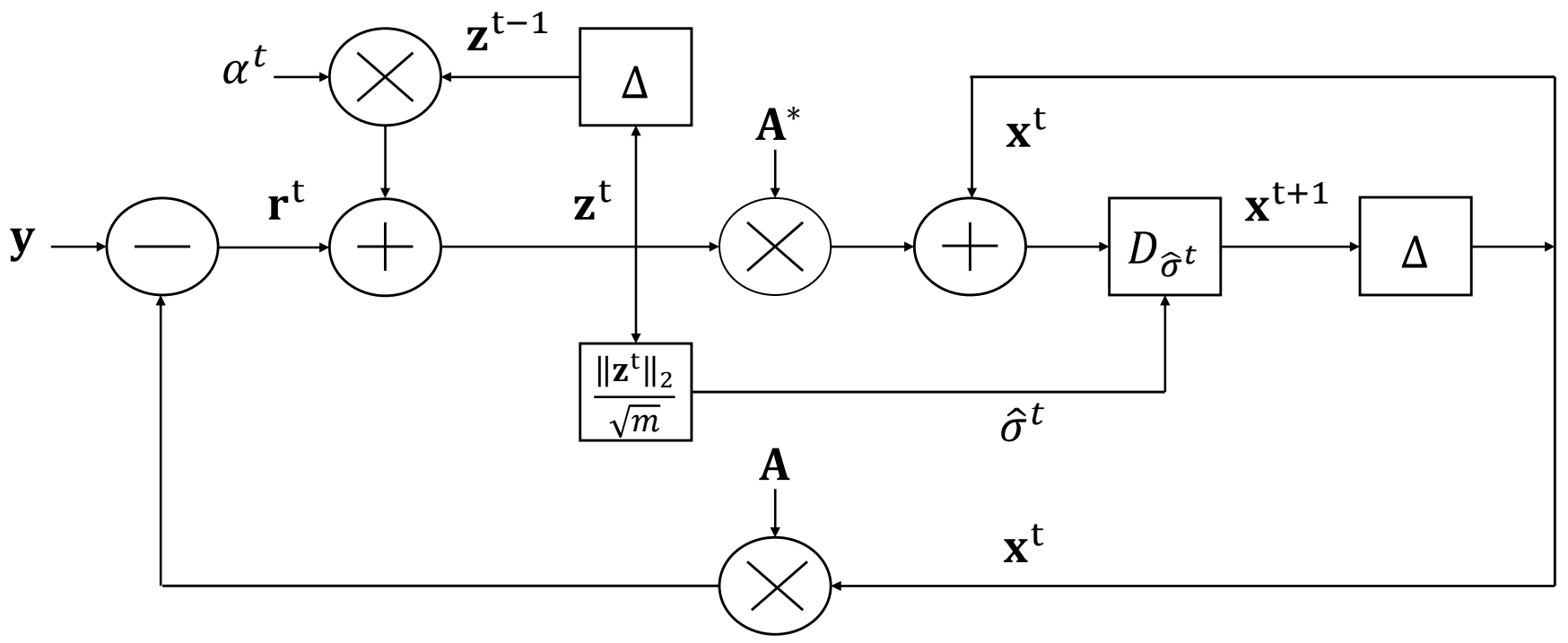

**FIGURE 6.** Generic block diagram for iterative denoising reconstruction algorithms: D-AMP, DAMP-D and IDA. This depicts equations 13, 14 and 15 in the text. The delay blocks $\Delta$ impart a one-iteration delay such that the output is the vector in the previous iteration. The Onsager term is $\alpha^t \mathbf{z^{t-1}}$ with $\alpha^t$ given by equation 24 and is derived from the divergence of the $\mathbf{x^t}$ vector from the previous iteration.

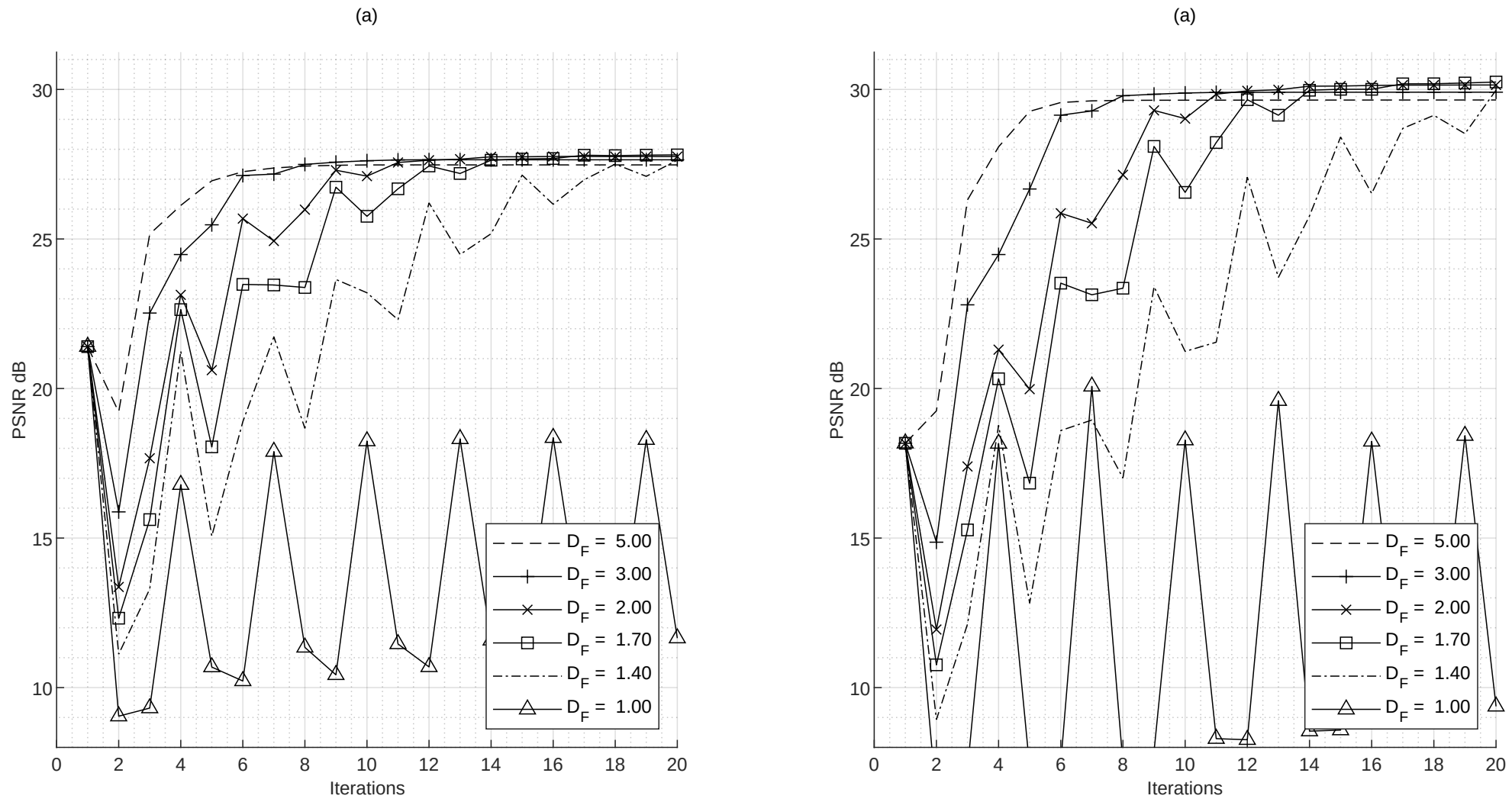

**FIGURE 7.** Reconstructing AL-DCT-DD using (a) D-AMP (when $D_F = 1.00$) and IDA (when $D_F > 1.00$) with a BM3D denoiser and (b) using D-AMP and IDA with a DnCNN denoiser. The results shown for the $256 \times 256$ Monarch image sensed at $C_R = 0.1$.

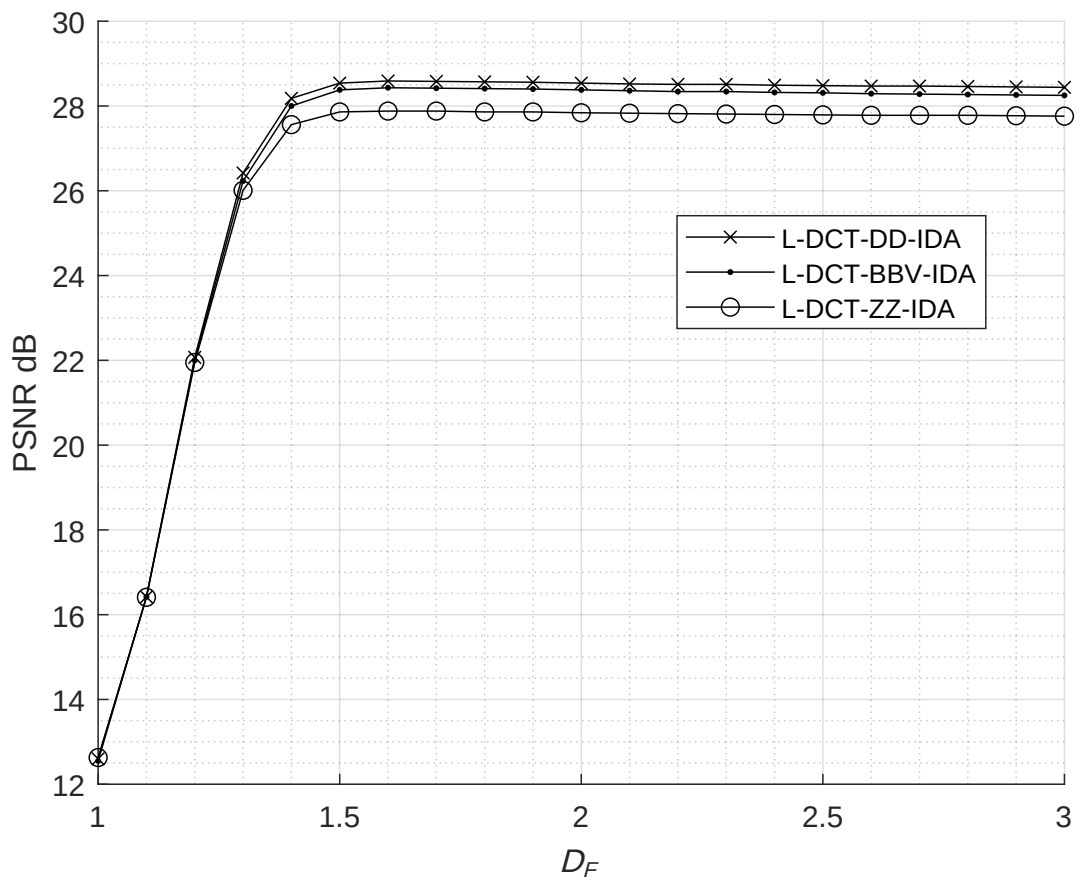

**FIGURE 8.** PSNR versus $D_F$ for L-DCT-ZZ, AL-DCT-BBV and AL-DCT-DD reconstructed using IDA for the $256 \times 256$ image set, with compression ratio $C_R = 0.1$.

by introducing a damping factor $D_F$. Then, $\alpha^t$ in equation (24) becomes:

$$\alpha^t = \frac{1}{\delta} \frac{div[D_{\hat{\sigma}^{t-1}}(\mathbf{x^{t-1}} + \mathbf{A}^* \mathbf{z^{t-1}})]}{m} \frac{1}{D_F} \tag{27}$$

The modified algorithm is referred to as D-AMP Damped (DAMP-D) with damping factor $D_F$.

### 2) Simplified D-AMP - IDA

Since the Onsager term requires the residual error to be Gaussian, it was hypothesized that a simplified version of the iterative denoising algorithm might provide better performance. Indeed, simplifying $\alpha^t$ to:

$$\alpha^t = \frac{1}{D_F} \tag{28}$$

was found to provide better performance than DAMP-D. The ensuing algorithm is called the IDA with damping factor $D_F$.

Figure 7 shows that IDA without damping, i.e., $D_F = 1$, did not reconstruct the AL-DCT-DD-sensed Monarch image from image set $256 \times 256$ satisfactorily using either of the denoisers. The PSNR decreased on the first iteration and then oscillated around a level that was 7.5 dB to 10 dB lower. However, when $D_F$ increased above 1, following the initial dip, the PSNR tended to increase. With $D_F \geq 1.4$, when employing the BM3D denoiser, IDA improved the initial AL-DCT-DD PSNR by approximately 2 dB, the maximum value after 20 iterations being reached with $D_F = 2$. When employing the DnCNN denoiser, IDA improved the AL-DCT-DD PSNR by 4.4 dB after 20 iterations. The results for the other images in the $256 \times 256$ image set are reported in supplementary [49] figure 9.

As IDA using the DnCNN denoiser is both substantially faster than with the BM3D denoiser (refer to table 9 in [49] for run-time comparison) and achieves superior performance, we present only results for this algorithm in this paper.

#### 3) Tuning $D_F$

To find the optimal value of $D_F$, the PSNR was plotted for $D_F$ between 1.0 and 3.0 for various algorithms, image sets and compression ratios. For example, figure 8 shows the PSNR versus $D_F$ plotted for L-DCT-ZZ, AL-DCT-BBV and AL-DCT-DD reconstructed using IDA for the $256 \times 256$ image set, with $C_R = 0.1$. $D_F$ = 2.0 was empirically found to be a good compromise value across the compression ratios of interest.

#### 4) BM3D and DnCNN Denoisers

Metzler et al. [42] implemented their D-AMP algorithm as an unrolled deep neural network, replacing the BM3D [22] with DnCNN [21], which is known to be a more accurate and faster denoiser. In this paper, we present the performance of iterative reconstruction using the DnCNN denoiser due to its superior performance.

The version of DnCNN used in this paper consists of 17 layers. The first layer consists of 64, 3x3x1 convolutional filters. The next fifteen layers consist of 64, 3x3x64 convolutional filters, followed by batch normalization [51] and a rectified linear unit (ReLU) [52]. The final layer consists of 3x3x64 convolutional filters.

## VII. EMPIRICAL INVESTIGATION

In this section, L-DCT-ZZ, AL-DCT-BBV and AL-DCT-DD are first compared empirically when reconstructed directly using the inverse 2D DCT (IDCT) and using CS reconstruction with the D-AMP and IDA algorithms on the $256 \times 256$ and $512 \times 512$ image sets. The CS algorithms were implemented using a DnCNN denoiser as detailed in section VI-4. The two adaptive algorithms were then compared with other published encoder-side and decoder-side adaptive CS algorithms, as well as with full sensing techniques. Finally, IDA was compared with two state-of-the-art non-adaptive algorithms, CREAM [19] and BCS-Net [20]. Since none of the techniques published in the literature were accompanied by source code, the method we used to compare with these published results was to repeat the simulation on the same image test sets and at the same compression ratios.

The simulations were executed on a server equipped with an Intel Xeon CPU E5-160 v3 clocked at 3.50 GHz, with 32.00 GB of RAM, running MATLAB version 2019a on Windows 10. The D-AMP algorithm was downloaded together with the D-AMP toolbox from [53]. The DnCNN code and models were downloaded from https://github.com/ricedsp/prDeep and require the MatConvNet package from https://www.vlfeat.org/matconvnet/.

### A. COMPARING THE DETERMINISTIC ALGORITHMS USING DIRECT AND CS RECONSTRUCTION

Table 3 presents the reconstruction results of the baseline linear L-DCT-ZZ algorithm and the two adaptive algorithms presented in this paper: AL-DCT-BBV and AL-DCT-DD. The results are also presented when using the original D-AMP algorithm and the IDA algorithm presented in this paper as a post-processing step.

When no CS reconstruction post-processing was utilised, AL-DCT-DD achieved the best results at all compression ratios $(0.01 - 0.5)$ on $256 \times 256$ images. PSNR performance improved by an average of 0.97 dB over the baseline L-DCT-ZZ algorithm across all compression ratios, and SSIM improved by an average of 0.0077. This improvement suffered no penalty in reconstruction execution time, as shown in table 8. At an image resolution of $512 \times 512$, this average performance increased in PSNR and SSIM by 0.79 dB and 0.0054, respectively. However, AL-DCT-BBV achieved a marginally higher PSNR increase for compression ratios in the range $0.1 - 0.5$ than AL-DCT-DD.

The original D-AMP as a post-processing reconstruction algorithm failed to reconstruct images at very low compression ratios $(0.01 - 0.04)$. This is as expected, as this algorithm was not designed to handle images that have been sensed using our deterministic sensing matrix. Interestingly, D-AMP achieved some success in reconstructing images at higher compression ratios, but the quality degraded in all cases.

IDA successfully improved the performances of all three algorithms. PSNR improved by an average of 2.04 dB and 0.83 dB across all compression ratios for $256 \times 256$ and $512 \times 512$ image resolutions, respectively. Similarly, SSIM improved by 0.0475 and 0.0229, respectively. It is interesting to note that AL-DCT-BBV-IDA achieved the best performance amongst the higher compression ratios $(0.10 - 0.50)$, while AL-DCT-DD-IDA was the best amongst the low compression ratios $(0.01 - 0.04)$. This has interesting implications in future schemes that utilise these algorithms in video transmission, mainly those that employ the group of picture structure [54] wherein key frames are sensed at a relatively high compression ratio and non-key frames at a substantially lower compression ratio.

| $C_R$ | 256x256 Image Set | | | | | | 512x512 Image Set | | | | | |
|---|---|---|---|---|---|---|---|---|---|---|---|---|
| | L-DCT-ZZ | | L-DCT-ZZ-D-AMP | | L-DCT-ZZ-IDA | | L-DCT-ZZ | | L-DCT-D-AMP | | L-DCT-IDA | |
| | PSNR dB | SSIM | PSNR dB | SSIM | PSNR dB | SSIM | PSNR dB | SSIM | PSNR dB | SSIM | PSNR dB | SSIM |
| **0.01** | 20.09 | 0.5187 | 4.82 | 0.0000 | 20.79 | 0.5831 | 22.85 | 0.5343 | 4.99 | 0.0000 | 23.53 | 0.5753 |
| **0.02** | 21.68 | 0.5955 | 5.86 | 0.0000 | 22.65 | 0.6739 | 24.23 | 0.5925 | 5.71 | 0.0049 | 24.97 | 0.6327 |
| **0.04** | 23.40 | 0.6806 | 6.71 | 0.1530 | 24.87 | 0.7687 | 25.70 | 0.6685 | 6.97 | 0.1738 | 26.63 | 0.7102 |
| **0.10** | 26.44 | 0.8096 | 19.84 | 0.7432 | 28.42 | 0.8733 | 28.63 | 0.7940 | 19.67 | 0.6853 | 29.60 | 0.8204 |
| **0.20** | 29.37 | 0.8889 | 28.89 | 0.8828 | 31.95 | 0.9281 | 31.62 | 0.8767 | 28.25 | 0.8162 | 32.63 | 0.8923 |
| **0.30** | 31.95 | 0.9308 | 31.06 | 0.9208 | 34.85 | 0.9563 | 33.82 | 0.9164 | 29.64 | 0.8629 | 34.74 | 0.9273 |
| **0.40** | 34.38 | 0.9562 | 31.96 | 0.9282 | 37.31 | 0.9725 | 35.90 | 0.9411 | 30.16 | 0.8680 | 36.68 | 0.9487 |
| **0.50** | 36.93 | 0.9728 | 33.41 | 0.9568 | 39.69 | **0.9823** | 38.06 | 0.9593 | 32.31 | 0.9094 | 38.73 | **0.9643** |
| **0.01 - 0.04** | 21.72 | 0.5983 | 5.80 | 0.0510 | 22.77 | 0.6752 | 24.26 | 0.5984 | 5.89 | 0.0596 | 25.04 | 0.6394 |
| **0.10 - 0.50** | 31.81 | 0.9117 | 29.03 | 0.8864 | 34.45 | 0.9425 | 33.61 | 0.8975 | 28.01 | 0.8284 | 34.47 | 0.9106 |
| $C_R$ | AL-DCT-BBV | | AL-DCT-BBV-D-AMP | | AL-DCT-BBV-IDA | | AL-DCT-BBV | | AL-DCT-BBV-D-AMP | | AL-DCT-BBV-IDA | |
| | PSNR dB | SSIM | PSNR dB | SSIM | PSNR dB | SSIM | PSNR dB | SSIM | PSNR dB | SSIM | PSNR dB | SSIM |
| **0.01** | 20.09 | 0.5187 | 4.82 | 0.0000 | 20.79 | 0.5831 | 22.85 | 0.5343 | 4.99 | 0.0000 | 23.53 | 0.5753 |
| **0.02** | 21.68 | 0.5955 | 5.86 | 0.0000 | 22.65 | 0.6739 | 24.23 | 0.5925 | 5.71 | 0.0049 | 24.97 | 0.6327 |
| **0.04** | 23.00 | 0.6629 | 6.71 | 0.1661 | 24.38 | 0.7436 | 25.74 | 0.6636 | 7.10 | 0.1292 | 26.74 | 0.7035 |
| **0.10** | 26.70 | 0.7982 | 15.63 | 0.6863 | 28.73 | 0.8672 | 29.18 | 0.7923 | 13.96 | 0.5735 | 30.30 | 0.8216 |
| **0.20** | 30.54 | 0.8930 | 29.36 | 0.8891 | **33.37** | **0.9359** | 32.72 | 0.8805 | 22.77 | 0.8033 | **33.85** | **0.8976** |
| **0.30** | 33.32 | 0.9334 | 30.93 | 0.9112 | **36.42** | **0.9607** | 35.26 | 0.9191 | 27.63 | 0.8533 | **36.32** | **0.9303** |
| **0.40** | 35.90 | 0.9561 | 33.15 | 0.9403 | **39.01** | **0.9733** | 37.36 | 0.9419 | 31.03 | 0.8860 | **38.25** | **0.9492** |
| **0.50** | 38.29 | 0.9702 | 33.88 | 0.9459 | **41.19** | 0.9810 | 39.36 | 0.9571 | 33.08 | 0.9080 | **40.02** | 0.9612 |
| **0.01 - 0.04** | 21.59 | 0.5924 | 5.79 | 0.0554 | 22.61 | 0.6669 | 24.27 | 0.5968 | 5.93 | 0.0447 | 25.08 | 0.6372 |
| **0.10 - 0.50** | 32.95 | 0.9102 | 28.59 | 0.8746 | **35.74** | 0.9436 | 34.78 | 0.8982 | 25.69 | 0.8048 | **35.75** | **0.9120** |
| $C_R$ | AL-DCT-DD | | AL-DCT-DD-D-AMP | | AL-DCT-DD-IDA | | AL-DCT-DD | | AL-DCT-DD-D-AMP | | AL-DCT-DD-IDA | |
| | PSNR dB | SSIM | PSNR dB | SSIM | PSNR dB | SSIM | PSNR dB | SSIM | PSNR dB | SSIM | PSNR dB | SSIM |
| **0.01** | 20.32 | 0.5271 | 5.32 | 0.0000 | **21.05** | **0.5918** | 23.09 | 0.5417 | 5.06 | 0.0000 | **23.81** | **0.5837** |
| **0.02** | 22.02 | 0.6085 | 5.64 | 0.0087 | **23.09** | **0.6880** | 24.60 | 0.6042 | 5.83 | 0.0019 | **25.35** | **0.6434** |
| **0.04** | 23.99 | 0.6998 | 7.76 | 0.2146 | **25.43** | **0.7810** | 26.33 | 0.6826 | 7.36 | 0.1784 | **27.13** | **0.7170** |
| **0.10** | 27.07 | 0.8173 | 20.38 | 0.7697 | **29.13** | **0.8784** | 29.40 | 0.8000 | 17.20 | 0.6771 | **30.31** | **0.8227** |
| **0.20** | 30.48 | 0.8953 | 29.92 | 0.8944 | 33.14 | 0.9308 | 32.50 | 0.8797 | 27.14 | 0.8156 | 33.39 | 0.8938 |
| **0.30** | 33.46 | 0.9350 | 31.19 | 0.9228 | 36.13 | 0.9563 | 34.94 | 0.9179 | 30.00 | 0.8556 | 35.67 | 0.9269 |
| **0.40** | 36.10 | 0.9588 | 32.67 | 0.9324 | 38.78 | 0.9722 | 37.08 | 0.9408 | 30.68 | 0.8741 | 37.70 | 0.9467 |
| **0.50** | 38.58 | 0.9735 | 33.72 | 0.9524 | 40.97 | 0.9811 | 39.19 | 0.9595 | 33.21 | 0.9083 | 39.77 | 0.9634 |
| **0.01 - 0.04** | 22.11 | 0.6118 | 6.24 | 0.0744 | **23.19** | **0.6869** | 24.67 | 0.6095 | 6.08 | 0.0601 | **25.43** | **0.6480** |
| **0.10 - 0.50** | 33.14 | 0.9160 | 29.57 | 0.8943 | 35.63 | **0.9438** | 34.62 | 0.8996 | 27.64 | 0.8261 | 35.37 | 0.9107 |

**TABLE 3.** L-DCT-ZZ, AL-DCT-BBV and AL-DCT-DD with IDCT, D-AMP and IDA reconstruction on the $256 \times 256$ and $512 \times 512$ image sets, using the DnCNN denoiser. Maximum PSNR and SSIM values for all compression ratios are in bold.

### B. COMPARISON WITH ADAPTIVE CS RESULTS IN THE LITERATURE

L-DCT-ZZ and the two adaptive algorithms achieved state-of-the art performance, in PSNR and SSIM terms, across many adaptive CS algorithms proposed in the literature. Table 4 summarizes the PSNR results, where our real-time algorithms perform better than the ES-CS, DS-CS and ES-FS BCS algorithms in the literature. Each row in table 4 reports the PSNR result of a published ABCS scheme and those of our algorithms on the same image set.

Only JRW-BCS, JRW-BCS-Sol2-EB and JRW-BCS-Sol2-VB from [36] and ABCS-SF-D [34] achieve better results than L-DCT-ZZ, with our AL-DCT-BBV and AL-DCT-DD algorithms achieving the best performance in all cases.

Table 5 compares AL-DCT-BBV and AL-DCT-DD, reconstructed using IDA with the DnCNN denoiser, against seven adaptive algorithms whose performance cannot be matched by our real-time algorithms. All our CS reconstructed algorithms outperformed StatACS and JRW-BCS-Sol2-SB in terms of PSNR. StatACS is not block-based, although it can be reconstructed in real time. AL-DCT-BBV-IDA outperformed the three InVDS algorithms in [36].

None of our CS reconstructed algorithms exceed the PSNR performance of the full sensing, Sol1-VB and Sol1-SB variants presented in [36]. A comparison with the full sensing THB and THI algorithms is fairer. In this case, THB achieved 39.03 dB and THI 40.87 dB, exceeding the best Sol1 variant by 3.55 dB and 5.39 dB, respectively.

### C. COMPARISON WITH TWO STATE-OF-THE-ART, NON-ADAPTIVE ALGORITHMS

In this section, we compare our algorithms with CREAM [19] and BCS-Net [20]. Table 6 shows that the real-time AL-DCT-DD algorithm achieved a better PSNR than CREAM at $C_R = 0.1$, but CREAM was superior to our real-time algorithms in all other cases. When our adaptive algorithms were reconstructed using IDA, the situation reversed, with both always performing better. The best PSNR and SSIM results were achieved by AL-DCT-BBV-IDA followed by AL-DCT-DD-IDA for $D_F = 2$ and with 15 iterations.

| Algorithm | Type | PSNR dB | L-DCT-ZZ | AL-DCT-BBV | AL-DCT-DD |
|---|---|---|---|---|---|
| | | | PSNR dB | PSNR dB | PSNR dB |
| ABCS-TVAL3 [38] | ES-CS | 30.82 | 37.38 | **39.35** | 38.68 |
| JRW-BCS [36] | ES-CS | 33.50 | 33.25 | **34.26** | 34.19 |
| ABCS-Zhang [40] | DS-CS | 29.21 | 30.06 | **33.73** | 31.21 |
| Proposed in [41] | DS-CS | 31.08 | 32.88 | 34.16 | **34.44** |
| JRW-BCS-Sol2-EB [36] | DS-CS | 34.08 | 33.25 | 34.19 | **34.26** |
| JRW-BCS-Sol2-VB [36] | DS-CS | 34.04 | 33.25 | 34.19 | **34.26** |
| ABCS-Wang [28] | ES-FS | 28.36 | 35.22 | **36.85** | 36.54 |
| Var-reg OP3 [29] | ES-FS | 29.18 | 30.39 | 31.02 | **31.64** |
| Proposed [32] | ES-FS | 27.08 | 31.48 | **32.93** | 32.41 |
| ABCS-Canh [33] | ES-FS | 32.16 | 34.37 | 35.59 | **35.62** |
| ABCS-SF-D [34] | ES-FS | 32.64 | 32.26 | **33.84** | 33.19 |
| JND [35] | ES-FS | 28.35 | 32.13 | 33.34 | **33.36** |

TABLE 4. Comparison of L-DCT-ZZ, AL-DCT-DD and AL-DCT-BBV with ABCS results in the literature derived on the image sets in the literature, using the compression ratios in the referenced publication. Best results are highlighted in bold. The underlined results are the only inferior L-DCT-ZZ results.

| Algorithm | Type | Image Set | [37] | [39] | [36] |
|---|---|---|---|---|---|
| | | RA | PSNR dB | PSNR dB | PSNR dB |
| StatACS [37] | ES-CS | IWT | 31.57 | | |
| InVDS-WT [39] | ES-CS | DAMP | | 34.72 | |
| InVDS-DCT [39] | ES-CS | DAMP | | 34.83 | |
| InVDS-HT [39] | ES-CS | DAMP | | 34.40 | |
| JRW-BCS-Sol1-EB [36] | ES-FS | NESTA | | | 34.59 |
| JRW-BCS-Sol1-VB [36] | ES-FS | NESTA | | | 35.41 |
| JRW-BCS-Sol1-SB [36] | ES-FS | NESTA | | | 35.48 |
| JRW-BCS-Sol2-SB [36] | DS-CS | NESTA | | | 34.53 |
| AL-DCT-BBV-IDA-DnCNN | ES-CS | IDA-DnCNN | **32.12** | **35.06** | 35.12 |
| AL-DCT-DD-IDA-DnCNN | ES-CS | IDA-DnCNN | 32.30 | 34.41 | 35.07 |
| L-DCT-THB | ES-FS | IDCT | | | 39.03 |
| L-DCT-THI | ES-FS | IDCT | | | **40.87** |

TABLE 5. Comparison of AL-DCT-BBV-IDA and AL-DCT-DD-IDA using the DnCNN denoiser with adaptive CS algorithms in the literature using the indicated reconstruction algorithm. The results in each PSNR column were derived using the image set in the cited publication, using the compression ratios in the referenced publication.

The results in table 6 were obtained with an average reconstruction time of 7.85 s over all compression ratios, with 15 iterations. L-DCT-ZZ, AL-DCT-BBV and AL-DCT-DD were reconstructed in fewer than 40 ms. AL-DCT-BBV-IDA reconstructed an image in 5.2 s on average, with 15 iterations at a compression ratio of 0.4.

Comparing our algorithms with BCS-Net, Table 7 shows that the best average PSNR and SSIM results were obtained by AL-DCT-DD-IDA with $D_F = 2$ and 15 iterations. AL-DCT-BBV-IDA achieves better results at higher compression ratios. The average reconstruction time over all compression ratios was approximately 3.5 s for 15 iterations. The reconstruction time for a compression ratio of 0.1 was approximately 2.37 s for 15 iterations.

### D. COMPLEXITY ANALYSIS

Block-based L-DCT-ZZ and AL-DCT-DD acquire $M$ measurements using $M \times B^2$ operations, whereas normal CS algorithms require $M \times (H \times W)$ operations. AL-DCT-BBV requires $M_{BBV} + (M - M_{BBV}) \times B^2$ operations. These analogue computations occur in the optical domain on single-pixel and multi-pixel cameras. The adaptive algorithms also need to compute the number of phase-2 measurements per block in the digital domain. These total $n_B \times n_S$ computations in the case of AL-DCT-BBV and $n_B \times B^2$ in the case of AL-DCT-DD.

The direct reconstruction of L-DCT-ZZ, AL-DCT-BBV and AL-DCT-DD using the inverse 2D DCT requires M inversions of a $B \times B$ matrix. Implemented as a separable inverse transform, 2D-IDCT requires $\mathcal{O}(B^2 \log B)$ operations. Therefore, IDCT reconstruction requires $\mathcal{O}(MB^2 \log B)$ operations.

With reference to figure 6 and equation (27), D-AMP has a computational complexity dominated by the three matrix multiplications and the denoiser computation and is of order $\mathcal{O}(MN^2 + 2NM^2) + \mathcal{O}(D_{\widehat{\sigma^t}})$, where $\mathcal{O}(D_{\widehat{\sigma^t}})$ is the complexity of the denoiser.

With reference to equation (27), DAMP-D has the same order of complexity because it merely divides the Onsager term by $\alpha$ every iteration. However, equation (28) shows that IDA avoids the computation of the Onsager term altogether and has complexity $\mathcal{O}(MN^2 + NM^2) + \mathcal{O}(D_{\widehat{\sigma^t}})$. Simulation results show that D-AMP and DAMP-D reconstruct a $256 \times 256$ image in approximately 2.4 s when using a DnCNN denoiser, whereas IDA-DnCNN requires 1.8 s. These results are in agreement with the theoretical complexity results.

On our platform, the L-DCT-ZZ, AL-DCT-BBV and AL-DCT-DD algorithms reconstruct $512 \times 512$ images in less than 30 ms. The improvement in performance with CS reconstruction comes at the expense of reconstruction time. Table 8 compares the reconstruction times of these algorithms with and without IDA post-processing using the DnCNN denoiser on the $256 \times 256$ image set.

## VIII. CONCLUSIONS AND FURTHER WORK

Two adaptive algorithms, AL-DCT-BBV and AL-DCT-DD, have been described that adapt the number of deterministic 2D-DCT measurements collected from block-based compressive image sensors. Both the BBV and the DD techniques proposed have been shown to estimate the number of transform coefficients per block that correlates well with the number of transform coefficients estimated by the full 2D DCT analysis. AL-DCT-BBV and AL-DCT-DD achieve state-of-the art performance in adaptive block CS of images that can be used in non-GPU-assisted, real-time applications, such as image and video capture with single-pixel cameras, multi-pixel cameras or focal plane processing image sensors. The algorithms reconstruct the compressively sensed images in real time, for example, less than 8 ms and 30 ms for $256 \times 256$ and $512 \times 512$ images, respectively.

| $C_R$ | 0.1 | | 0.2 | | 0.3 | | 0.4 | | Average | |
|---|---|---|---|---|---|---|---|---|---|---|
| **Algorithm** | **PSNR dB** | **SSIM** | **PSNR dB** | **SSIM** | **PSNR dB** | **SSIM** | **PSNR dB** | **SSIM** | **PSNR dB** | **SSIM** |
| CREAM | 27.31 | 0.8206 | 31.67 | 0.9014 | 34.14 | 0.9318 | 36.11 | 0.9510 | 32.31 | 0.9012 |
| L-DCT-ZZ | 26.92 | 0.8030 | 29.67 | 0.8795 | 31.82 | 0.9182 | 33.71 | 0.9429 | 30.53 | 0.8859 |
| AL-DCT-BBV | 27.28 | 0.7982 | 30.64 | 0.8823 | 33.34 | 0.9226 | 35.57 | 0.9459 | 31.71 | 0.8873 |
| AL-DCT-DD | 27.67 | 0.8131 | 30.78 | 0.8865 | 33.19 | 0.9223 | 35.22 | 0.9446 | 31.72 | 0.8916 |
| L-DCT-ZZ-IDA | 28.41 | 0.8506 | 31.61 | 0.9105 | 33.82 | 0.9385 | 35.70 | 0.9571 | 32.39 | 0.9142 |
| AL-DCT-BBV-IDA | 28.85 | 0.8486 | **32.73** | **0.9158** | **35.43** | **0.9433** | **37.48** | **0.9591** | **33.62** | **0.9167** |
| AL-DCT-DD-IDA | **29.28** | **0.8575** | 32.68 | 0.9128 | 34.89 | 0.9385 | 36.76 | 0.9554 | 33.40 | 0.9161 |

**TABLE 6.** Comparing CREAM PSNR and SSIM with L-DCT-ZZ, AL-DCT-BBV and AL-DCT-DD reconstructed using IDA with the DnCNN denoiser, with $D_F = 2.0$ and 15 iterations. Maximum PSNR/SSIM values are in bold and second largest underlined.

| $C_R$ | 0.01 | | 0.03 | | 0.05 | | 0.1 | | 0.2 | | 0.3 | | 0.4 | | Average | |
|---|---|---|---|---|---|---|---|---|---|---|---|---|---|---|---|---|
| **Algorithm** | **PSNR** | **SSIM** | **PSNR** | **SSIM** | **PSNR** | **SSIM** | **PSNR** | **SSIM** | **PSNR** | **SSIM** | **PSNR** | **SSIM** | **PSNR** | **SSIM** | **PSNR** | **SSIM** |
| BCS-Net | 20.88 | 0.5505 | 24.47 | 0.7087 | 26.04 | 0.7723 | 29.43 | 0.8676 | 33.06 | 0.9283 | 35.60 | 0.9554 | 36.70 | 0.9662 | 29.45 | 0.8213 |
| L-DCT-ZZ | 20.33 | 0.5072 | 22.82 | 0.6369 | 24.60 | 0.7260 | 27.06 | 0.8193 | 30.17 | 0.8953 | 32.70 | 0.9340 | 35.01 | 0.9569 | 27.53 | 0.7822 |
| AL-DCT-BBV | 20.33 | 0.5072 | 22.91 | 0.6424 | 24.17 | 0.7031 | 27.29 | 0.8112 | 31.08 | 0.8980 | 33.90 | 0.9363 | 36.29 | 0.9563 | 28.00 | 0.7792 |
| AL-DCT-DD | 20.56 | 0.5163 | 23.27 | 0.6519 | 25.03 | 0.7366 | 27.68 | 0.8272 | 31.16 | 0.9013 | 34.06 | 0.9376 | 36.53 | 0.9591 | 28.33 | 0.7900 |
| L-DCT-ZZ-IDA | 20.99 | 0.5615 | 24.09 | 0.7153 | 25.81 | 0.7864 | 28.78 | 0.8721 | 32.29 | 0.9275 | 34.90 | 0.9544 | 37.05 | **0.9698** | 29.13 | 0.8267 |
| AL-DCT-BBV-IDA | 20.99 | 0.5615 | 24.09 | 0.7153 | 25.58 | 0.7676 | 29.09 | 0.8676 | **33.29** | **0.9316** | **36.11** | **0.9567** | **38.35** | 0.9694 | 29.64 | 0.8242 |
| AL-DCT-DD-IDA | **21.25** | **0.5717** | **24.48** | **0.7240** | **26.30** | **0.7934** | **29.45** | **0.8769** | 33.27 | 0.9296 | 35.95 | 0.9543 | 38.23 | 0.9694 | **29.85** | **0.8313** |

**TABLE 7.** Comparing BCS-Net PSNR and SSIM with L-DCT-ZZ, AL-DCT-BBV and AL-DCT-DD reconstructed using IDA with the DnCNN denoiser, with $D_F = 2$ and 15 iterations. Maximum PSNR values are in bold and second largest underlined.

| Algorithm | L-DCT-ZZ | AL-DCT-BBV | AL-DCT-DD | L-DCT-ZZ-IDA | AL-DCT-BBV-IDA | AL-DCT-DD-IDA |
|---|---|---|---|---|---|---|
| $C_R$ | Time s | Time s | Time s | Time s | Time s | Time s |
| 0.01 | 0.0069 | 0.0069 | 0.0067 | 0.69 | 0.69 | 0.69 |
| 0.02 | 0.0080 | 0.0074 | 0.0070 | 0.79 | 0.79 | 0.80 |
| 0.04 | 0.0067 | 0.0069 | 0.0066 | 1.05 | 1.02 | 1.07 |
| 0.10 | 0.0067 | 0.0071 | 0.0067 | 1.86 | 1.75 | 1.84 |
| 0.20 | 0.0067 | 0.0068 | 0.0066 | 3.20 | 2.89 | 3.11 |
| 0.30 | 0.0067 | 0.0069 | 0.0065 | 4.42 | 4.09 | 4.38 |
| 0.40 | 0.0068 | 0.0069 | 0.0067 | 5.89 | 5.38 | 5.72 |
| 0.50 | 0.0067 | 0.0067 | 0.0066 | 7.86 | 6.50 | 7.08 |

**TABLE 8.** Comparison of reconstruction times for L-DCT-ZZ, AL-DCT-BBV and AL-DCT-DD using 2D-IDCT and IDA reconstruction with the DnCNN denoiser on the $256 \times 256$ image set, $D_F = 2$ and 20 iterations.

The original D-AMP algorithm frequently fails to reconstruct or improve our algorithms using deterministic sensing matrices. A block diagram representation of the D-AMP algorithm has been derived and used to interpret D-AMP as a feedback system with integral control. A modified version, DAMP-D, which damps the Onsager term with a factor $D_F$ and a simplified version, called the iterative denoising algorithm (IDA), has been derived that performs better when reconstructing our deterministic CS algorithms. The IDA algorithm simplifies the Onsager term to $\mathbf{z}^{\mathbf{t}}/D_F$ and achieves a considerable reduction in computational complexity and better results than D-AMP and DAMP-D. Both modified algorithms can be used as a post-processing technique on the AL-DCT-BBV and AL-DCT-DD algorithms. Setting $D_F = 2.0$ has been empirically found to be close to optimal for the $256 \times 256$ and $512 \times 512$ image sets used in our simulations.

The two adaptive algorithms reconstructed using IDA with a DnCNN denoiser outperform two state-of-the-art algorithms in terms of PSNR and SSIM. AL-DCT-BBV-IDA outperforms the GPU-assisted CREAM PSNR and SSIM by 1.31 dB and 0.0155, respectively. AL-DCT-DD-IDA outperforms the DNN BCS-Net PSNR and SSIM by 0.40 dB and 0.01, respectively. Even the non-CS reconstructed, real-time algorithms perform close to CREAM and BCS-Net, with low to moderate PSNR penalties, for example, within 0.59 dB of the CREAM PSNR and within 1.12 dB of BCS-Net, making them suitable for real-time compressive image and video sensing.

We also note that by utilising DCT measurements, the proposed algorithms can leverage the well-established source and channel coding techniques developed for hybrid DPCM/DCT encoding standards. The AL-DCT-BBV algorithm has better performance at $C_R$ values greater than 0.1, with AL-DCT-DD performing better at lower $C_R$ values. This means that in video coding applications, the AL-DCT-BBV algorithm will be more suitable for high-quality intra-coded frames and AL-DCT-DD for the more compressed predicted frames.

The improvement in performance by damping and simplifying the Onsager term was examined empirically. The development of a theoretical framework to exploit the interpretation of IDA as feedback systems with an integral component is of interest for further work.

## REFERENCES

[1] S. Qaisar, R. M. Bilal, W. Iqbal, M. Naureen, and S. Lee, "Compressive sensing: From theory to applications, a survey," Journal of Communications and Networks, vol. 15, no. 5, pp. 443–456, 2013.

[2] Wei Chen, Y. Andreopoulos, I. J. Wassell, and M. R. D. Rodrigues, "Towards energy neutrality in energy harvesting wireless sensor networks: A case for distributed compressive sensing?" in 2013 IEEE Global Communications Conference (GLOBECOM), 2013, pp. 474–479.

[3] D. Cao, K. Yu, S. Zhuo, Y. Hu, and Z. Wang, "On the implementation of compressive sensing on wireless sensor network," in 2016 IEEE First International Conference on Internet-of-Things Design and Implementation (IoTDI), 2016, pp. 229–234.

[4] E. J. Candes, J. Romberg, and T. Tao, "Robust uncertainty principles: exact signal reconstruction from highly incomplete frequency information," IEEE Transactions on Information Theory, vol. 52, no. 2, pp. 489–509, Feb 2006.

[5] D. L. Donoho, "Compressed sensing," IEEE Transactions on Information Theory, vol. 52, no. 4, pp. 1289–1306, April 2006.

[6] M. F. Duarte, M. A. Davenport, D. Takhar, J. N. Laska, T. Sun, K. F. Kelly, and R. G. Baraniuk, "Single-pixel imaging via compressive sampling," IEEE Signal Processing Magazine, vol. 25, no. 2, pp. 83–91, March 2008.

[7] X. Yuan, G. Huang, H. Jiang, and P. A. Wilford, "Block-wise lensless compressive camera," in 2017 IEEE International Conference on Image Processing (ICIP), pp. 31–35, 2017.

[8] R. Robucci, J. D. Gray, L. K. Chiu, J. Romberg, and P. Hasler, "Compressive sensing on a CMOS separable-transform image sensor," Proceedings of the IEEE, vol. 98, no. 6, pp. 1089–1101, 2010.

[9] E. J. Candes and T. Tao, "Decoding by linear programming," IEEE transactions on information theory, vol. 51, no. 12, pp. 4203–4215, 2005.

[10] S. G. Mallat and Z. Zhang, "Matching pursuits with time-frequency dictionaries," IEEE Transactions on signal processing, vol. 41, no. 12, pp. 3397–3415, 1993.

[11] D. Baron, S. Sarvotham, and R. G. Baraniuk, "Bayesian compressive sensing via belief propagation," IEEE Transactions on Signal Processing, vol. 58, no. 1, pp. 269–280, 2009.

[12] D. L. Donoho, A. Maleki, and A. Montanari, "Message-passing algorithms for compressed sensing," Proceedings of the National Academy of Sciences, vol. 106, no. 45, pp. 18 914–18 919, 2009.

[13] C. A. Metzler, A. Maleki, and R. G. Baraniuk, "From denoising to compressed sensing," IEEE Transactions on Information Theory, vol. 62, no. 9, pp. 5117–5144, 2016.

[14] A. Mousavi, A. B. Patel, and R. G. Baraniuk, "A deep learning approach to structured signal recovery," in 2015 53rd Annual Allerton Conference on Communication, Control, and Computing (Allerton). IEEE, pp. 1336–1343, 2015.

[15] Lu Gan, "Block compressed sensing of natural images," in 2007 15th International Conference on Digital Signal Processing, pp. 403–406, July 2007.

[16] S. Mun and J. E. Fowler, "Block compressed sensing of images using directional transforms," in 2010 Data Compression Conference, pp. 547–547, March 2010.

[17] J. E. Fowler, S. Mun, E. W. Tramel et al., "Block-based compressed sensing of images and video," Foundations and Trends in Signal Processing, vol. 4, no. 4, pp. 297–416, 2012.

[18] Zhou Wang, A. C. Bovik, H. R. Sheikh, and E. P. Simoncelli, "Image quality assessment: from error visibility to structural similarity," IEEE Transactions on Image Processing, vol. 13, no. 4, pp. 600–612, April 2004.

[19] C. Zhao, J. Zhang, R. Wang, and W. Gao, "Cream: Cnn-regularized admm framework for compressive-sensed image reconstruction," IEEE Access, vol. 6, pp. 76 838–76 853, 2018.

[20] S. Zhou, Y. He, Y. Liu, and C. Li, "Multi-channel deep networks for block-based image compressive sensing," arXiv preprint arXiv:1908.11221, 2019.

[21] K. Zhang, W. Zuo, Y. Chen, D. Meng, and L. Zhang, "Beyond a gaussian denoiser: Residual learning of deep cnn for image denoising," IEEE Transactions on Image Processing, vol. 26, no. 7, pp. 3142–3155, 2017.

[22] K. Dabov, A. Foi, V. Katkovnik, and K. Egiazarian, "Image denoising by sparse 3-d transform-domain collaborative filtering," IEEE Transactions on image processing, vol. 16, no. 8, pp. 2080–2095, 2007.

[23] C. Chen, E. W. Tramel, and J. E. Fowler, "Compressed-sensing recovery of images and video using multihypothesis predictions," in 2011 Conference Record of the Forty Fifth Asilomar Conference on Signals, Systems and Computers (ASILOMAR), pp. 1193–1198, Nov 2011.

[24] E. W. Tramel and J. E. Fowler, "Video compressed sensing with multihypothesis," in 2011 Data Compression Conference. IEEE, 2011, pp. 193–202.

[25] J. Romberg, "Imaging via compressive sampling," IEEE Signal Processing Magazine, vol. 25, no. 2, pp. 14–20, 2008.

[26] I. IEC, "Information technology-digital compression and coding of continuous-tone still images: Requirements and guidelines," Standard, ISO IEC, pp. 10 918–1, 1994.

[27] X. Yuan and R. Haimi-Cohen, "Image compression based on compressive sensing: End-to-end comparison with jpeg," arXiv preprint arXiv:1706.01000, 2017.

[28] A. Wang, L. Liu, B. Zeng, and H. Bai, "Progressive image coding based on an adaptive block compressed sensing," IEICE Electronics Express, vol. 8, no. 8, pp. 575–581, 2011.

[29] W. Guicquero, A. Dupret, and P. Vandergheynst, "An adaptive compressive sensing with side information," in 2013 Asilomar Conference on Signals, Systems and Computers, pp. 138–142, Nov 2013.

[30] Z. Gao, C. Xiong, L. Ding, and C. Zhou, "Image representation using block compressive sensing for compression applications," Journal of Visual Communication and Image Representation, vol. 24, no. 7, pp. 885–894, 2013.

[31] H. Zheng and X. Zhu, "Sampling adaptive block compressed sensing reconstruction algorithm for images based on edge detection," The Journal of China Universities of Posts and Telecommunications, vol. 20, no. 3, pp. 97–103, 2013.

[32] J. Luo, Q. Huang, S. Chang, and H. Wang, "Fast reconstruction with adaptive sampling in block compressed imaging," IEICE Electronics Express, vol. 11, no. 6, pp. 20 140 056–20 140 056, 2014.

[33] T. Nguyen Canh, "Adaptive block compressive sensing: Toward a complete approach," 10.13140/RG.2.1.1807.4728, 2015.

[34] W. Wang, W. Yang, and J. Li, "An adaptive sampling method of compressed sensing based on texture feature," Optik-International Journal for Light and Electron Optics, vol. 127, no. 2, pp. 648–654, 2016.

[35] J. Wu, Y. Wang, K. Zhu, and Y. Zhu, "Perceptual sparse representation for compressed sensing of image," in 2016 Visual Communications and Image Processing (VCIP). IEEE, pp. 1–4, 2016.

[36] S. Zhu, B. Zeng, and M. Gabbouj, "Adaptive sampling for compressed sensing based image compression," Journal of Visual Communication and Image Representation, vol. 30, pp. 94–105, 2015.

[37] A. Averbuch, S. Dekel, and S. Deutsch, "Adaptive compressed image sensing using dictionaries," SIAM Journal on Imaging Sciences, vol. 5, no. 1, pp. 57–89, 2012.

[38] Nguyen et al, "Rate allocation method for block-based compression sensing," Journal of Broadcasting Engineering, vol. 20, no. 3, pp. 398–407, 05 2015.

[39] J. Liu and C. Ling, "Adaptive compressed sensing using intra-scale variable density sampling," IEEE Sensors Journal, vol. 18, no. 2, pp. 547–558, Jan 2018.

[40] X. Zhang, A. Wang, B. Zeng, L. Liu, and Z. Liu, "Adaptive block-wise compressive image sensing based on visual perception," IEICE TRANSACTIONS on Information and Systems, vol. 96, no. 2, pp. 383–386, 2013.

[41] A. Akbari, D. Mandache, M. Trocan, and B. Granado, "Adaptive saliency-based compressive sensing image reconstruction," in 2016 IEEE International Conference on Multimedia Expo Workshops (ICMEW), July, pp. 1–6, 2016.

[42] C. Metzler, A. Mousavi, and R. Baraniuk, "Learned d-amp: Principled neural network based compressive image recovery," in Advances in Neural Information Processing Systems, pp. 1772–1783, 2017.

[43] B. Shi, Q. Lian, and S. Chen, "Compressed sensing magnetic resonance imaging based on dictionary updating and block-matching and three-dimensional filtering regularisation," IET Image Processing, vol. 10, no. 1, pp. 68–79, 2016.

[44] B. Shi, Q. Lian, X. Huang, and N. An, "Constrained phase retrieval: when alternating projection meets regularization," JOSA B, vol. 35, no. 6, pp. 1271–1281, 2018.

[45] B. Shi, Q. Lian, and H. Chang, "Deep prior-based sparse representation model for diffraction imaging: A plug-and-play method," Signal Processing, vol. 168, p. 107350, 2020.

[46] S. Foucart and H. Rauhut, "A mathematical introduction to compressive sensing," Bull. Am. Math, vol. 54, pp. 151–165, 2017.

[47] A. L. Pilastri and J. M. R. Tavares, “Reconstruction algorithms in compressive sensing: an overview,” in 11th edition of the Doctoral Symposium in Informatics Engineering (DSIE-16), 2016.

[48] I. Daubechies, M. Defrise, and C. De Mol, “An iterative thresholding algorithm for linear inverse problems with a sparsity constraint,” Communications on Pure and Applied Mathematics: A Journal Issued by the Courant Institute of Mathematical Sciences, vol. 57, no. 11, pp. 1413–1457, 2004.

[49] “Supplementary material and data for: Adaptive block compressive imaging: towards a real-time and low complexity implementation,” available at: http://dx.doi.org/10.21227/r2t1-s503, 2020.

[50] N. Ahmed, T. Natarajan, and K. R. Rao, “Discrete cosine transform,” IEEE transactions on Computers, vol. 100, no. 1, pp. 90–93, 1974.

[51] S. Ioffe and C. Szegedy, “Batch normalization: Accelerating deep network training by reducing internal covariate shift,” arXiv preprint arXiv:1502.03167, 2015.

[52] A. Krizhevsky, I. Sutskever, and G. E. Hinton, “Imagenet classification with deep convolutional neural networks,” in Advances in neural information processing systems, pp. 1097–1105. 2012.

[53] “D-amp toolbox,” original-date: 2017-02-09T18:27:40Z., available at https://github.com/ricedsp/D-AMP_Toolbox., Apr. 2017.

[54] Z. Krusevac and Z. Bojkovic, “Bit allocation and coding gain in mpeg video compression standard for optimum picture ordering,” in Fifth Asia-Pacific Conference on ... and Fourth Optoelectronics and Communications Conference on Communications,, vol. 2, 1999, pp. 927–930 vol.2.

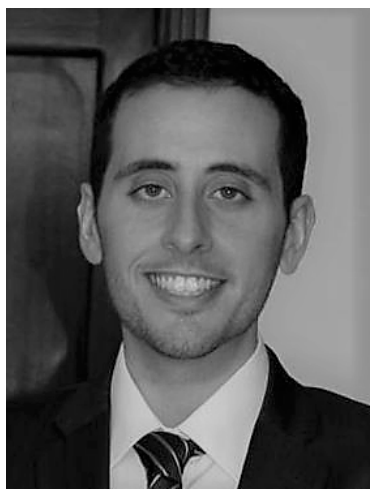

JOSEPH ZAMMIT received his B.Sc. degree in communications and computer engineering from the University of Malta, Msida, Malta, in 2012, his M.Sc. degree in wireless and optical communications, from University College London, London, U.K., in 2013 and is a Ph.D. candidate at the Computer Laboratory, University of Cambridge, Cambridge, U.K. His research interests include sparse representations, wireless networks, deep learning, and distributed systems.

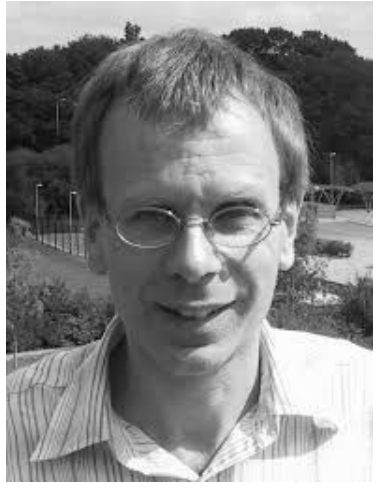

IAN J. WASSELL received his B.Sc. and B.Eng. degrees from the University of Loughborough, Loughborough, U.K., in 1983, and the Ph.D. degree from the University of Southampton, Southampton, U.K., in 1990. He is a Senior Lecturer in the Computer Laboratory, University of Cambridge, Cambridge, U.K. He has in excess of 15 years of experience in the simulation and design of radio communication systems gained via a number of positions in industry and higher education. He has published more than 180 papers concerning wireless communication systems. His research interests include fixed wireless access, sensor networks, cooperative networks, propagation modelling, compressive sensing, and cognitive radio. He is a member of the IET and a Chartered Engineer.